\documentclass[twocolumn,epsfig, graphics,floatfix]{revtex4-1}

\usepackage{amsmath,amsfonts,amssymb,graphics,graphicx,epsfig,color,times}
\usepackage{mathtools}
\usepackage{amsthm}
\usepackage{psfrag}
\usepackage{braket}
\usepackage{graphicx}
\usepackage{color,soul}
\AtBeginDocument{\usepackage{booktabs}}
\usepackage[hidelinks]{hyperref}
\hypersetup{colorlinks=false}
\usepackage{float}
\usepackage{enumitem}
\usepackage{multirow}

\usepackage{array}
\newcolumntype{L}{>{\arraybackslash}m{6.5cm}}

\usepackage[framemethod=tikz]{mdframed}
\usepackage{hyperref}

\theoremstyle{plain}

\theoremstyle{plain}

\theoremstyle{definition}

\theoremstyle{remark}

\usepackage{xcolor}

\usepackage[normalem]{ulem}
\usepackage{dsfont}

\begin{document}

\title{Macromux: scalable postselection for high-threshold fault-tolerant quantum computation}

\author{Patrick Birchall, Jacob Bridgeman, Christopher Dawson, Terry Farrelly, Yehua Liu, Naomi Nickerson, Mihir Pant, Sam Roberts, Karthik Seetharam, David Tuckett}

\affiliation{PsiQuantum}
\date\today

\begin{abstract}
We introduce a new resource-efficient scheme for fault-tolerant quantum computation known as `macroscale multiplexing' (or simply `Macromux'), that utilizes \textit{scalable postselection} to significantly improve the threshold of a given fault-tolerant protocol against both Pauli and erasure errors. 
Macromux is a hierarchical method for postselecting on constant-size space-time windows of a fault tolerant protocol, requiring only constant additional overheads. The method can be straightforwardly implemented for any fault-tolerant protocol and in any architecture that has access to routing and memory, such as linear-optical fusion-based architectures. 
We construct fault-tolerant protocols that, to our knowledge, have the highest thresholds in the literature; 
we perform simulations of fusion-based schemes based on the surface code, showing a maximum possible increase in Pauli thresholds of up to a factor of $\sim6$ (from $1.0\%$ to $5.9\%$). 
Our schemes are highly-resource efficient, and can for example, double the loss thresholds of some photonic fusion-based protocols using as little as $3 \times$ overhead. 

\end{abstract}

\maketitle

\section{Introduction} 

The path toward large-scale universal quantum computation is based on fault tolerance. 
To perform a computation fault tolerantly, we pay the price in overhead in terms of both space and time, and these overheads are prohibitive for current devices. 
In recent years, however, there have been rapid developments in the theory of fault tolerance, with the discovery of new quantum error-correcting codes and protocols that require increasingly lower overheads~\cite{panteleev2021degenerate, breuckmann2021balanced, panteleev2022asymptotically, bravyi2024high, tamiya2024polylog, yamasaki2024time, litinski2025blocklet,yoder2025tour}, as well as in the quality of hardware to realise them~\cite{bluvstein2024logical,paetznick2024demonstration,reichardt2024demonstration,google2025quantum,PsiHardware,larsen2025integrated}.
Despite these staggering advances, constructing fault-tolerant protocols with high thresholds remains an important problem, as many quantum computing architectures have physical error rates that are too high to take advantage of these low-overhead schemes~\cite{endote1}.

\begin{figure}
	\includegraphics[width=0.47\textwidth]{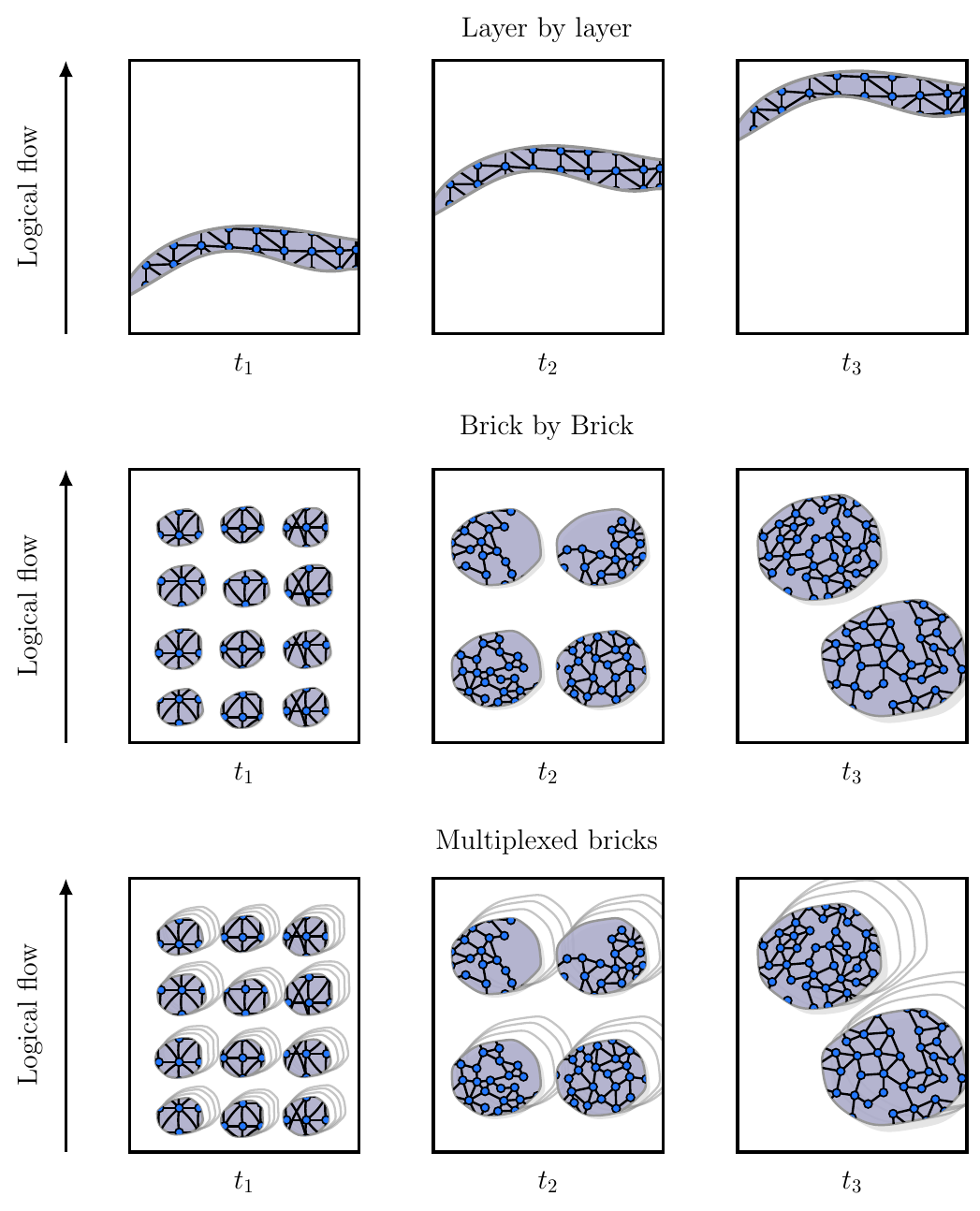} 
	\caption{The idea behind Macromux. A fault-tolerant computation consists of a series of operations on a set of qubits in space-time. In fusion-based quantum computation, these blue nodes correspond to small entangled states called resource states and black edges are fusions (Bell-basis measurements). (top row) Standard fault-tolerant constructions perform operations layer-by-layer, at times $t_1$, $t_2$, $t_3$, consistent with the flow of logical information e.g. following a quantum circuit. (middle row) For stabilizer-based fault-tolerance, we can modify the order of instructions; instead of performing operations layer by layer, we may perform them in parallel in disjoint space-time ``bricks''. For fusion-based quantum computation, each brick consists of a constant number of resource states and fusions. (bottom row) By partitioning the fault-tolerant computation into bricks of constant size, one may create multiple copies of each brick in parallel. Bricks can be grouped together to form larger bricks (e.g., to go from $t_1$ to $t_2$), in a way that favourably shapes the error configuration. By repeating this process up to some scale and keeping only the highest-quality bricks at the final stage, the computation can be performed with lower logical error rates.}
	\label{figConceptual}
\end{figure}

In this paper, we introduce \textit{macromux}, an approach to construct high threshold fault-tolerant protocols using multiplexing methods. In its simplest form, multiplexing is a method whereby multiple attempts at generating a target state are performed. This is related to postselection, a method of increasing interest to reduce error rates in quantum error correcting codes~\cite{akahoshi2025runtime, lee2025efficient, chen2025scalable, english2024thresholds, paetznick2024demonstration, xie2026simple}. A barrier preventing the use of many of these postselection methods in realistic computations is scalability; if the postselection acceptance rate decreases with the code distance or computation size, then it is unable to be effectively used in a full fault-tolerant algorithm. 

Macromux solves the scalability problem by utilizing a hierarchical multiplexing approach to fault tolerance that divides a fault tolerant protocol into constant-size windows called ``bricks''.
Multiple copies of each brick are generated, ranked based on their error quality, and combined in a hierarchical way to produce bigger bricks, as schematically shown in Fig.~\ref{figConceptual}. 
In doing so, errors can be filtered or favourably combined, leading to increased error tolerances at larger scales, allowing a direct increase in the error threshold of a protocol with a tunable multiplexing overhead (which can be viewed analogously to a ``postselection overhead''). 
This method can be applied to any kind of fault tolerant protocol, including topological~\cite{kitaev1997quantum, dennis2002topological,bombin2006topological}, concatenated~\cite{knill1996concatenated, aharonov1997fault, aliferis2005quantum, Knill05, yamasaki2024time, yoshida2025concatenate, litinski2025blocklet}, and LDPC~\cite{tillich2014quantum, panteleev2021asymptotically, breuckmann2021ldpc, breuckmann2021balanced, bravyi2024high, jacob2025single}. Closely related to macromux is Knill's celebrated scheme for fault tolerance \cite{Knill05}, which makes use of postselection in concatenated codes to get ultra high circuit-level thresholds. Macromux provides us with a general prescription for constructing high threshold schemes, while focussing on resource efficiency. 

At each stage, the quality of each brick is ranked by a function called the ``scorer". 
The scorer takes information about errors that have occurred in a brick, such as losses/erasures of qubits/outcomes as well as syndromes if they are available.  Macromux can remove both erasure and Pauli errors. 

We consider two methods for scoring when doing postselection. 
The first is a simpler scorer based on counting syndromes and erasures. 
The second is a soft-information scorer, which we call the frozen gap scorer. 
This uses the logical gap to score error configurations, which was proposed in \cite{bombin2023logical}, building on techniques introduced in \cite{hutter2014efficient}, and studied in several other works~\cite{pattison2023hierarchical,smith2024mitigating, gidney2023yoked,meister2024efficient,gidney2024magic}.

\textbf{Macromux in fusion-based quantum computation.}
We present our scheme in terms of the fusion-based model of quantum computation (FBQC) \cite{bartolucci2023fusion}, but the ideas can be applied to any model of computation and architecture, such as conventional circuit-based computations, provided one has access to a suitable qubit memory and qubit routing, which are natural hardware primitives for fusion-based quantum computation with linear optics.
In linear-optical fusion-based quantum computation, the computer produces a stream of constant-size resource states: small entangled states, which are fused together to form fault-tolerant logical blocks~\cite{bombin2023logical}.
In macromux, each brick is comprised of a constant number of resource states and fusions between them. 
Macromux is therefore a natural extension of the multiplexing that occurs during resource state generation -- which is used to overcome the probabilistic nature of entanglement in linear optical circuits -- as bricks can be viewed as larger resource states.
A key feature of macromux is the ability to exploit the fault tolerance of the protocol, i.e., the checks and the Tanner/syndrome graph structure, which allows us to keep overheads low by tolerating imperfections (rather than trying to eliminate them) in the bricks by arranging errors into benign configurations. 

\begin{figure*}
	\centering
	\includegraphics[width=0.9\textwidth]{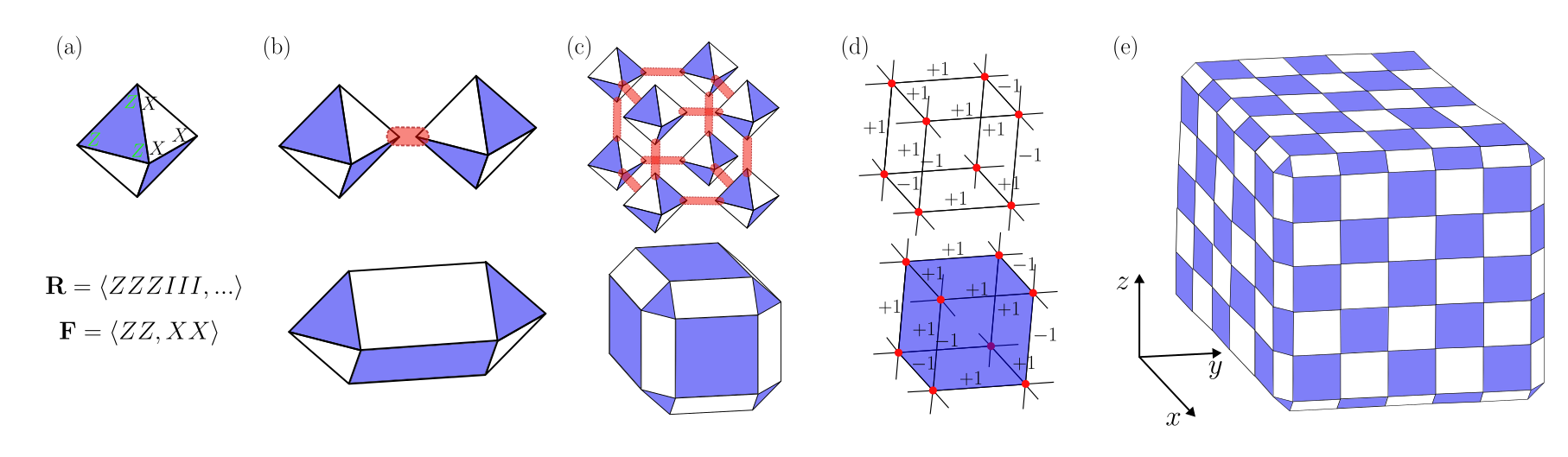} 
	\caption{Fusion-based quantum computation with surface-code resource states.  (This example is equivalent to the 6-ring fusion network of \cite{bartolucci2023fusion} up to Hadamards.)
    (a) shows a six-qubit resource state where each vertex is a qubit (one of the qubits is at the back of the octahedron).  Each blue face represents an $X$ stabilizer and each white face represents a $Z$ stabilizer, and faces sharing an edge always have the opposite colour.  Topologically this is a sphere.  The stabilizer resource state group $\mathcal{R}$ and the fusion measurement group $\mathcal{F}$ are both also shown.  (b) illustrates how a fusion between two qubits of two resource states creates a bigger surface-code resource state.  This can be understood by noting how measuring, e.g., the $ZZ$ operator merges $Z$-type stabilizers on the two resource states.  The qubits that are measured are now in a product state with the rest of the qubits and may be discarded (in linear-optical fusion-based quantum computation, the qubits would actually be destructively measured).  (c) continues the process, illustrating how fusing qubits from eight resource states arranged as shown creates a bigger surface-code state. This sequence of fusions generates a check operator, given by the product of the $ZZ$ measurement outcomes, with an example shown in (d).  In (e), we see the result of fusing many such resource states in this way: we get alternating $XX$ and $ZZ$ checks. 
    (Each check only overlaps on a single fusion measurement outcome, so the decoding problem is two syndrome graph decoding problems.)  By fusing qubits on pairs of opposite faces, we get periodic boundary conditions.  So if we fuse qubits on the two faces perpendicular to the x direction and do the same for the z direction, we get two toric codes on the faces perpendicular to the y direction.  One can show that these two toric codes are entangled giving us two logical Bell pairs.  (An alternative is to create two entangled planar codes by doing single-qubit measurements of the qubits on the x and z planes.)}
	\label{figConceptualOverview}
\end{figure*}

While general, this idea is especially intuitive in the topological code setting, where we use topologically-encoded resource states (surface codes defined on a topological sphere). 
By fusing these resource states, we build larger surface code states. 
By multiplexing these larger states, the locations of errors, which are normally distributed in 3D space-time, are mostly constrained to 2D surfaces, and are therefore easier to correct. 

\textbf{Fault tolerance with high thresholds.}
As a precise application, we present a family of (fault-tolerant) fusion networks whose Tanner graph (which describes the relationship between errors and the checks that detect them) under macromux approaches that of the idealized 2D Tanner graph of the surface code in the code capacity setting (i.e., the setting where stabilizer measurements are free of error) as we increase the macromux parameter. 
This gives an effective dimension reduction for the Tanner graph, increasing the Pauli threshold and erasure threshold by a factor of ~6x and ~5x, respectively. 
We emphasize that this is for a full fusion-based error model, which is the fusion-based quantum computation analogue of the circuit-based error model. 
Moreover, our schemes are highly-resource efficient. 
We give an example where macromux increases the threshold by a factor of $\sim 2$ with footprint increasing by a factor of $\sim 10$.

\section{Background: Fusion-based quantum computation}
A fusion-based quantum computation~\cite{bartolucci2023fusion} is defined by a \textbf{fusion network}: a set of resource states (constant-sized stabilizer states) with collective stabilizer group $\mathbf{R}$ and entangling fusion measurements between them described by a Pauli subgroup $\mathbf{F}$.
Here, for simplicity, we consider two assumptions on $\mathbf{R}$ and $\mathbf{F}$. 
\begin{itemize}
    \item[(i)] We consider Fusion measurements $f_{i,j}$ that are Bell-basis measurements involving two qubits $i,j$, such that $f_{i,j}$ is a measurement of $\{X_iX_j,Z_iZ_j\}$, giving $\mathbf{F} = \langle -I, X_iX_j,Z_iZ_j ~|~ {i,j}\rangle$ as the group of all measurements with signs.
    \item[(ii)] We consider fusion networks embeddable in 3-dimensions where resource states are 2D surface-code states on (topological) spheres~\cite{FusionComplexes}.
\end{itemize}
Such fusion networks are called \textit{surface-code fusion networks}, and let us represent the fusion network by a 3D graph where vertices correspond to resource states and edges represent the fusion measurements between them, and one dimension can be thought of as the logical time-like direction. That a graph is enough to uniquely specify both $\mathbf{F}$ and $\mathbf{R}$ is a property of surface-code networks as explained in~\cite{FusionComplexes}). However, the ideas here can all be extended to more general fusion networks. Fig.~\ref{figConceptualOverview} provides an example of a fusion network created from surface-code resource states.
Readers interested in the equivalence between FBQC and other models of quantum computation can utilize the ZX calculus~\cite{coecke2008interacting}, as described in~\cite{Bombin2024unifyingflavorsof}.

\textbf{Checks.}
Fusion networks achieve fault tolerance by generating checks, which comprise products of measurement outcomes that, in the absence of errors, have a known value (e.g., the product may be $+1$). 
These arise due to redundancies of elements in the fusion group $\mathbf{R}$ and the resource state stabilizers $\mathbf{S}$, and are described by the check group $\mathbf{C} = \mathbf{R} \cap \mathbf{F}$.
Because checks have a fixed outcome in the absence of errors, they can be used to detect the presence of errors.

\textbf{Syndrome and Tanner graphs.}
We can express the fault tolerant properties of the fusion network in terms of a Tanner graph. Given a generating set of $checks$, we place a check node for each generator, an outcome node for each measured outcome of $\mathbf{F}$, and connect outcome nodes to check nodes they are supported in. 
For surface-code fusion networks, there is a basis for the checks such that each measured outcome belongs to at-most two check generators, meaning all outcome nodes are degree two. 
As such we may define the \textit{syndrome graph} by removing outcome nodes from the Tanner graph and placing edges between check nodes whenever the checks share an outcome.   

\textbf{Linear optics \& resource-state generation.}
In a discrete-variable photonic architecture, resource states are fixed-size entangled states of photons using, e.g., the dual-rail encoding to represent qubits. Linear-optical fusions are probabilistic in nature, meaning the desired two-qubit outcomes are not always measured (even in the absence of photon loss or other errors), an event which is called \textit{fusion failure}. For example, for Type-II Bell state fusions~\cite{BR04}, successful measurement of the Bell basis $\{XX, ZZ\}$ occurs only $50\%$ of the time, while projection onto a product basis such as $\{XI, IX\}$ occurs the rest of the time. See, e.g., \cite{bartolucci2023fusion} for more details. Furthermore, single-photon sources and circuits for generating (small) entangled states of photons are also probabilistic.  

Because of the probabilistic nature of entanglement generation, a key component of photonic fusion-based quantum computation is multiplexing (muxing). 
In this context, multiplexing means to attempt a probabilistic operation (such as generating a state) multiple times and keep only the successful cases to increase the chance of success. 
The multiplexing is achieved with a switching network~\cite{bartolucci2021switch}, that allows photons of successful events to be separated from those in failed events.

At the smallest scale, multiple probabilistic heralded single-photon sources can be multiplexed to have reliable, on-demand single photon generation~\cite{PsiHardware}. 
This muxing process is continued at further stages.  For example, single photons are fed into probabilistic circuits that generate small seed states, such as Bell states and 3GHZ states. 
Again, to increase the probability of successful seed state generation, we multiplex by implementing several such circuits to ensure high enough probability of successful seed state generation. 
To make the larger resource states (such as the six-qubit state described in Fig.~\ref{figConceptualOverview}), we fuse these seeds states together. 
Since fusion measurements can fail (or return nonsense outcomes due to photon loss), again we must multiplex. 
(For more details on hardware for linear-optical photonic quantum computation, see Ref.~\cite{PsiHardware}.) 

Macromux is a natural extension of this style of muxing to the larger, fault-tolerant scale, where check structure and syndrome information can be used to mitigate and remove Pauli errors (as well as erasure errors due to fusion failure and loss).

\section{Macromux}

Defining a macromux protocol involves specifying (i) a
base fusion network, (ii) a dicing scheme, (iii) a macromux parameter $M$, and (iv) a scoring method.

\subsection{Dicing}

Given a base fusion network defined by $(\mathbf{R}, \mathbf{F})$, the first step in defining a macromux scheme is to partition the fusion network into disjoint pieces called bricks. 
We can do this without modifying the logical action or fault-tolerant properties of the network as all fusions commute and thus can be performed in any order.
Each brick is a connected region of the fusion graph, consisting of several resource states and fusions between them. 
The \textit{dicing} defines this partition across several stages in a hierarchical fashion. 
At stage one, each brick is an individual resource state. At stage two, each brick may be several resource states and fusions between them, partitioning the set of all resource states. Bricks at each stage are defined as a set of previous-stage bricks, along with fusions between them. 
The final stage consists of all resource states and remaining fusions defining the logical block.

\begin{figure}
	\centering
	\includegraphics[width=0.95\columnwidth]{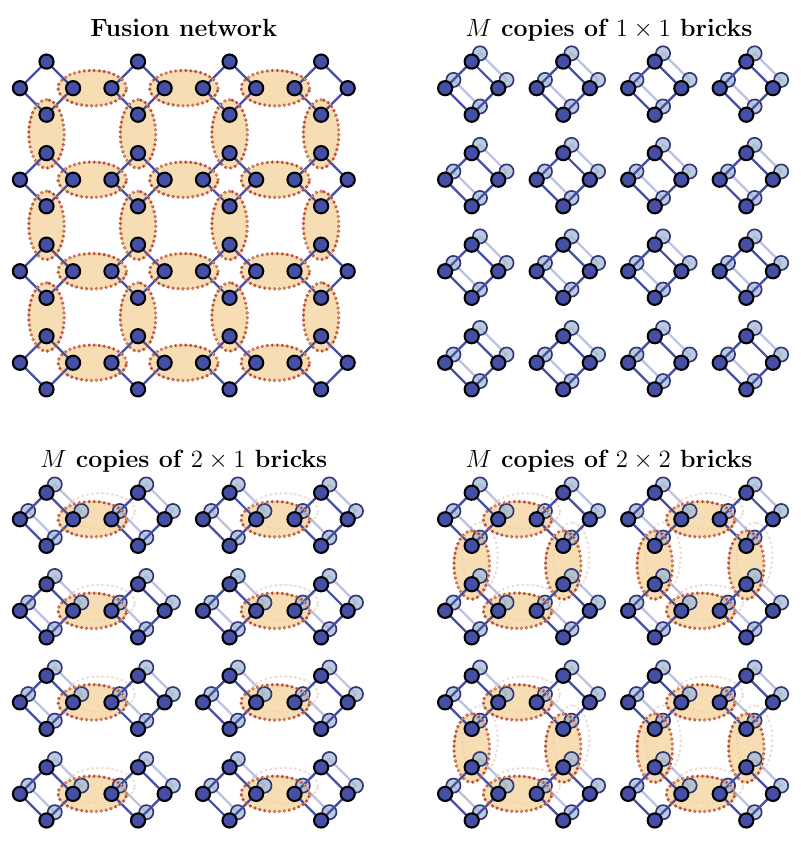}
	\caption{(Toy 2D example of a fusion network, diced into four stages. (top left) The fusion network: resource states are 4 qubit rings, and fusions are $\{XX,ZZ\}$ Bell measurements. Boundary conditions not shown (assume periodic, for example).
	(top right) The initial stage has bricks that are single resource states, and there are $M=2$ copies of each brick. 
	(bottom left) The second stage has bricks that are pairs of resource states fused together, and there are $M=2$ copies of each brick. 
	(bottom right) The third stage has bricks that are pairs of stage-2 bricks, and there are $M=2$ copies of each brick. The final stage fuses all bricks together to give the original fusion network.
	}
	\label{fig2DExample}
\end{figure}

\textbf{Dicing a 2D toy example}
Fig.~\ref{fig2DExample} shows a toy 2D network to illustrate this idea. 
The fusion network consists of 4-qubit resource states arranged on the vertices of a square graph. 
We dice the network into four stages, with the first stage having 1x1 bricks (which are individual resource states), the second stage having 2x1 bricks (two resource states), the third stage being 2x2 bricks (four resource states), and the fourth stage being the final fusion network (i.e. all bricks and the remaining fusions between them), as shown in Fig.~\ref{fig2DExample}.

\textbf{Dicing a 3D example: the 6-ring fusion network.}
For the 6-ring fusion network (see Fig.~\ref{figConceptualOverview}), a natural hierarchical scheme involves defining $N$ stages, where the bricks in each stage consist of pairs of bricks of the previous stage fused together in a cuboidal fashion. 
The first stage bricks are single resource states (called 1x1x1 bricks), the second stage comprises pairs of resource states fused together (2x1x1 bricks) and so on, up to a max brick size of ${n_1 \times n_2 \times n_3}$. 
The 2x1x1 and 2x2x2 bricks are shown in parts (b) and (c) of Fig.~\ref{figConceptualOverview}. 
After the chosen max brick size is achieved, all bricks are fused together in the final stage.

\textbf{Spherical macromux.}
For surface-code based fusion networks~\cite{FusionComplexes} such as the 6-ring fusion network, we call this scheme \textit{spherical macromux}, as at every stage, each resource state and each brick is (by definition) a non-degenerate surface code state on a topological sphere. 

\subsection{Overheads: Macromux parameter}
At each stage of macromux, the fusion network is divided into disjoint bricks, and we make $M$ copies of each of these bricks, where $M$ is called the macromux parameter. 
To form the brick copies at stage $i+1$, we group smaller bricks from stage $i$. 
So we must start with $M$ copies of each resource state at stage one, which means the overhead cost for the scheme is a factor of $M$ compared with the un-maxromuxed scheme. 

\subsection{Scoring}
\begin{figure}
	\centering
	\includegraphics[width=0.95\columnwidth]{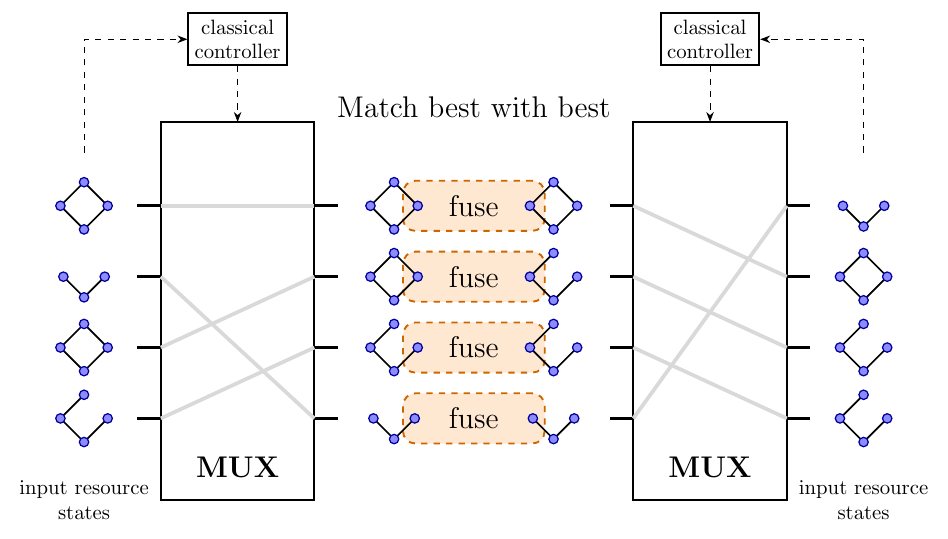} 
	\caption{(a) Schematic of how 1x1 bricks are ranked and fused together to create 2x1 bricks in the 2D example.  Here we have the possibility of imperfect resource states:\ the complete four-qubit ring state is the ideal, whereas other states with, e.g., cut edges or missing nodes are imperfect. Note that the routing schematic is for illustrative purposes only, with more efficient approaches possible~\cite{bartolucci2021switch}.}
	\label{figShuffling}
\end{figure}

Once a dicing scheme has been specified and a macromux parameter chosen, we need a method of determining which copies of each brick are to be fused together. 
We do this using a scorer, with the convention that higher scores correspond to higher quality bricks.  This allows us to rank the bricks by quality, so that bricks of equal rank are fused together
(i.e.,  we pair best with best, second-best with second-best, and so on). 
This is schematically illustrated in Fig.~\ref{figShuffling}.

Bricks can include checks, so syndrome information can be used by the scorer.
For example, in the 6-ring dicing scheme, this occurs for brick sizes 2x2x2 (see Fig.~\ref{figConceptualOverview}) and beyond.
This syndrome information is powerful, as it allows us to reduce the number of Pauli errors.  To be precise, the scorer is a function from erasure and syndrome configurations of a brick to a real number. 

\subsubsection{Count scorer}
Our first scorer is called the count scorer, and it  counts the number of erasures and syndrome bits in a brick, and combines them to a scalar
\begin{equation}
S_{\text{count}} = -\sum_{G\in\mathcal{G}}\exp\left(\alpha ||\vec{e}_G||_1 + \beta ||\vec{c}_G||_1\right).
\end{equation}
Here $\alpha\geq 0$ and $\beta\geq 0$ are free parameters that we tune to get the best performance.  The sum is over $\mathcal{G}$ the set of syndrome graphs of the brick.  $||\cdot||_1$ is the one norm, i.e., the sum of the absolute value of a vector's entries.  The vector $\vec{e}_G$ is the erasure vector for syndrome graph $G$, whose entries are $1$ if the corresponding edge is erased and $0$ otherwise.  The vector $\vec{c}_G$ is the syndrome after merging graph vertices incident on an erased edge.  (Merging checks in this way is also part of a method to decode in the presence of bit flips and erasures \cite{stace2009thresholds}.)  Note that it only makes sense to count syndrome bits from complete checks, i.e., those that have all incident outcomes measured.  The convention is that higher scores correspond to better bricks, which is the reason for the overall minus sign.  

A more sophisticated scorer is explained in the next section.

\subsubsection{Frozen-gap scorer}
More efficient scoring methods can be obtained by utilizing the structure of the error configuration. 
For instance, in surface-code protocols with large bricks, a brick copy with two well-separated syndromes is likely lower quality than one with two nearby syndromes because the former most likely has more Pauli errors -- however the count scorer would assign them the same score. 

Here we leverage a soft-output decoder to evaluate the quality of a brick by estimating the likelihood of a spanning error across the axes of a brick. 
This scorer builds upon the `logical gap' method of Ref.~\cite{bombin2024fault}, which was also developed in  Refs.~\cite{smith2024mitigating,gidney2023yoked,meister2024efficient}. 
However, an additional complication in our setting is that we have incomplete check information -- any qubits on the surface of a brick are yet to be fused, and so checks involving them may not yet be complete.
We introduce the \textit{frozen gap} as one method to perform soft-output decoding in the presence of only partial information. 
We first describe the logical gap, which works with complete syndrome information, before describing the frozen gap, which works with partial information.

\textbf{The logical gap.}
When a complete set of checks have been measured, the logical gap $\Delta$ is calculated by considering two minimum-weight recoveries, corresponding to inequivalent logical sectors, and taking the absolute value of the difference in their weights. 
For planar code schemes, we can efficiently do this using a slight modification of the minimum-weight perfect matching algorithm~\cite{bombin2024fault}. 
The logical gap is a measure of the confidence that the decoder is correct (a larger gap is better). 
See Fig.~\ref{figGapCalculation2} for an example calculation on a 2D syndrome graph, and App.~\ref{appFG} for how to evaluate the logical gap for general codes and protocols.

\begin{figure}
	\centering
    \includegraphics[width=0.99\columnwidth]{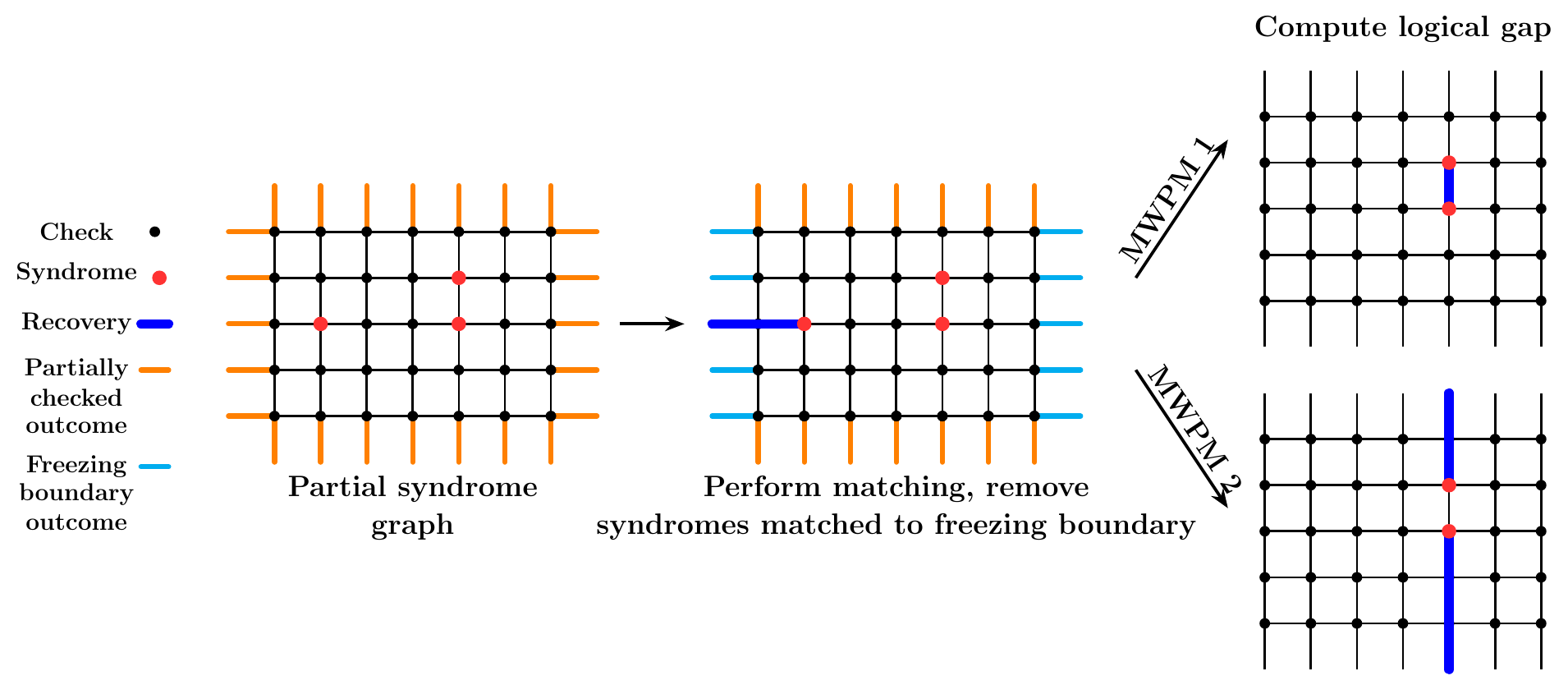} 
	\caption{Example computing the frozen gap on a 2D surface code syndrome graph with partially checked outcomes around the boundary. Nodes are syndrome locations and edges are error locations. We consider the top and bottom rough edges to be the gap boundary, while the left and right are freezing boundaries. We first perform a matching to potentially freeze out some syndromes, in this case the freezing weight is $w=2$. Then the logical gap is computed as $\Delta = |1-5| = 4$. The frozen gap is $\Delta_f = \max(4-2\phi, 0)$.}
	\label{figGapCalculation}
\end{figure}

\textbf{The frozen gap.}
When only a subset of the check operators are complete, then we can calculate the frozen gap. 
Consider the set of `partially checked outcomes', these are outcomes that belong to an incomplete (i.e., not yet measured) check. 
We partition these partially checked outcomes into two types of boundaries we call \textit{gap} and \textit{freezing} boundaries. 
The gap boundaries will be those between which we will measure the logical gap, while the freezing boundaries will be those which will be used to initially freeze out some syndromes. 

The frozen gap is computed as follows. 
First, we perform an initial decoding, where syndromes can be matched to each other as well as gap or freezing boundaries. 
Any syndromes matched to a freezing boundary are removed from the configuration, and the combined weight to match any syndromes to a freezing boundary is recorded as $w$ -- called the freezing weight. 
Next, with the updated syndrome configuration, the logical gap is computed as usual. 
The frozen gap is given by $\Delta_f(i) = \max(\Delta(i) - \phi w_i, 0)$, where $\phi \geq 0$ is a free parameter we tune to optimize performance; $\Delta(i)$ is the gap along the given axis $i$ after freezing out syndrome bits; and $w_i$ is the weight of the correction used to freeze out syndrome bits. 
This is outlined in Fig.~\ref{figGapCalculation}.

\textbf{Scoring with the frozen gap.}
For spherical macromux, there will be unchecked outcomes around the entire boundary of the brick. 
For each brick, we compute six frozen gaps, by considering gap-boundaries on antipodal sides of the brick, along each axis ($x$, $y$, and $z$) for both $X$ and $Z$-type syndrome graphs, with the remainder being freezing boundaries. 
See Fig.~\ref{figGapCalculation2} in App.~\ref{appFG} for more details on the calculation.

The score using the frozen gap is
\begin{equation}
    S_{\text{gap}} = -\sum_{i\in\mathrm{axes}}e^{-\delta \Delta_f(i)},
\end{equation}
where $i \in \mathrm{axes}$ are the total of six different axes for the $X$ and $Z$-type syndrome graphs; $\delta$ and $\phi$ are free parameters that we tune for best performance; $\Delta_f(i)$ is the frozen gap along the $i$ axis.

\subsection{Macromux extensions}

\textbf{Contextual scoring.}
Scoring need not be a function of a singular brick only -- it can also consider the additional environmental information of the bricks it is to be fused with.
A scorer that uses such environmental information is called \textit{contextual}.
For simplicity, we consider non-contextual scorers in the following.

\begin{figure}
	\centering
	\includegraphics[width=0.8\columnwidth]{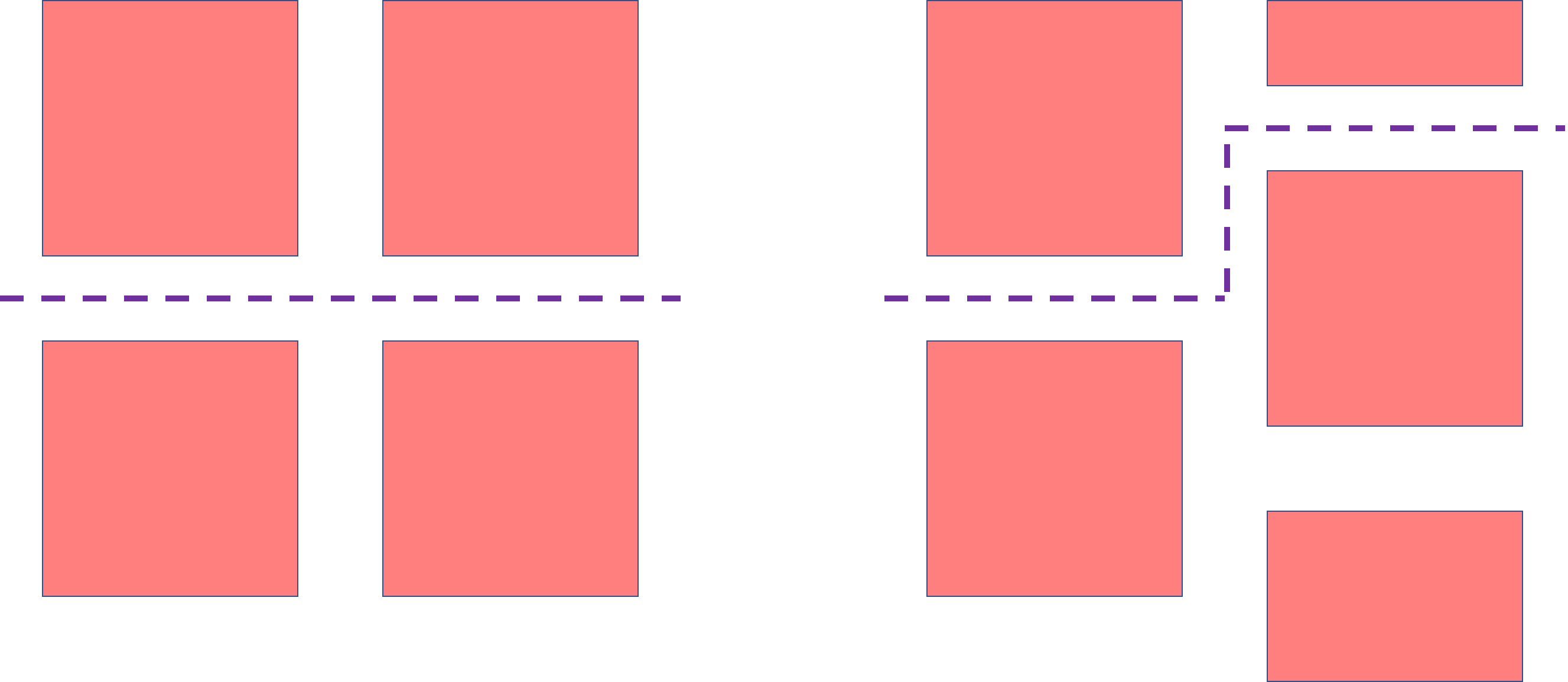} 
	\caption{A schematic 2D slice of the fusion network with bricks represented by red squares, comparing (left) no brick offsetting with (right) brick offsetting. Since macromux postselects out some Pauli errors and erasures, the most likely path for harmful error strings is between bricks. Offsetting can be used make this path longer.}
	\label{figOffsetting}
\end{figure}

\textbf{Offsetting}
In the above, we have defined translationally invariant dicings. However, when defining a dicing scheme, one may consider the use of \textit{offsetting}, which modifies the brick locations. 
This is particularly useful for certain choices of dicing where bricks may naively only include one type of check (e.g. $X$-type or $Z$-type only), or there is otherwise an asymmetry between the frequency of $X$ and $Z$ checks in bricks. 
By shifting the brick locations, offsetting can be used to balance the ratio of $X$ and $Z$ checks that belong to bricks and are therefore multiplexed. 
Offsetting can also be used to increase the effective distance of the protocol (see Figure \ref{figOffsetting}), by increasing the distance that error chains must traverse on un-muxed fusions (or in other words, minimal error paths must travel through bricks where they may be detected).

For example, in the 6-ring fusion network with a translationally invariant dicing into 2x2x2 bricks, each brick will only contain the one type of check (e.g., an $ZZ$ check as shown in Fig.~\ref{figConceptualOverview}). In this case, we will only detect (and reduce) Pauli errors that flip $X$ checks. 
This leads to an asymmetry in $X$ and $Z$ logical error rates and thresholds. 
As we are ultimately limited by the weaker of the two logical error rates, it is beneficial to keep the $X$ and $Z$ decoding problems symmetric. 
To remedy this, we consider offsetting bricks in different layers by translating them one lattice spacing in a direction within the layer. 
See Fig.~\ref{figResultsScheme} for a schematic illustration.

We note that in the presence of a biased error model, we may actually offset (or not) to deliberately imbalance the two decoding problems. 

\subsection{Physical realizations, routing \& switching, and beyond fusion-based quantum computation}
We have presented macromux in the context of FBQC, where a fault-tolerant protocol is expressed in terms of resource states and fusion. 
However, macromux can be utilized in any other model of quantum computation, including circuit-based and measurement-based quantum computation, for example. 
To do so, one must partition the fault-tolerant protocol -- which may be expressed as a quantum circuit, or ZX diagram~\cite{coecke2008interacting,van2020zx,Bombin2024unifyingflavorsof} -- into constant-size space-time windows called bricks, and operations between these bricks.
Each brick consists of a set of qubit preparations, gates and measurements. The dicing defines these brick operations, as well as the gates to be performed between the bricks (after the scoring and ordering has taken place).
Implementing a dicing scheme in general requires qubit routing and qubit memory, which are natural primitives for photonic (where bricks can be interleaved~\cite{bombin2021interleaving}), neutral atom, and trapped ion architectures, for example.

\section{Results}

\begin{figure}
	\centering
	\includegraphics[width=0.67\columnwidth]{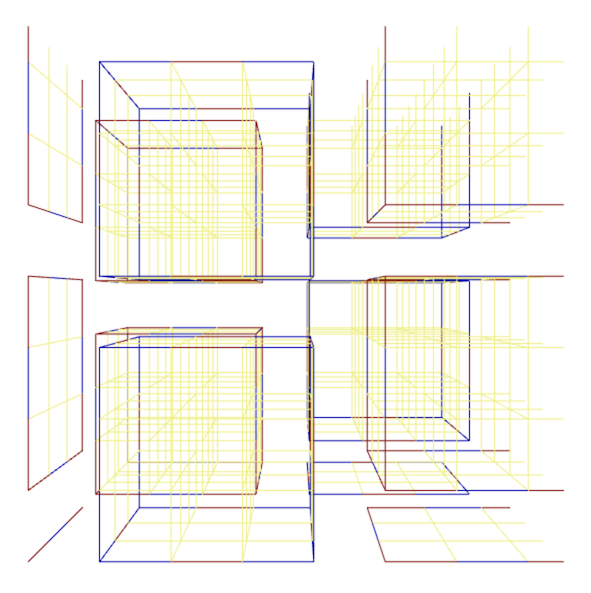} 
	\caption{A choice of offsetting for 4x4x4 bricks of the 6-ring network that maintains primal-dual balance. The fusion network is a cubic lattice. Yellow shaded edges are fusions where both outcomes belong to a primal or dual check within a brick. Blue (red) edges are fusions where only the primal (dual) outcome is checked within the brick.  Note the periodic boundary conditions mean that some bricks are split in the figure.}
	\label{figResultsScheme}
\end{figure}

We benchmark the performance of macromux in the setting of the 6-ring fusion network, for various brick sizes, macromux overheads, and scorers, under both erasure and Pauli noise. We consider the cuboidal diced scheme that includes offsetting, such that the $X$ and $Z$ decoding problems are symmetric, as shown in Fig.~\ref{figResultsScheme}.

We first focus on the max achievable threshold limit in this setting, where both bricks and macromux overheads are large, giving the highest known thresholds to our knowledge. 
We then show how to efficiently achieve extremely high thresholds with finite size bricks and finite macromux overheads. Finally, we provide loss-tolerance results for the 6-ring fusion network and loopy diamond network~\cite{FusionComplexes, bartolucci2025comparison} with a \textbf{cuboctahedral} dicing scheme as defined in Fig.~\ref{figbbs} under a range of local encodings. 

\subsection{Threshold limits with large brick sizes: approaching the 2D code capacity limit}
In Ref.~\cite{FusionComplexes} it was observed that fusion networks with higher thresholds typically require larger resource states. 
We can understand the converse of this result in the context of macromux: where resource states of larger size can be viewed as large bricks with many macromux copies, naturally giving higher thresholds from suppression of physical errors. 

\begin{figure}[h]
	\centering
	\includegraphics[width=0.95\columnwidth]{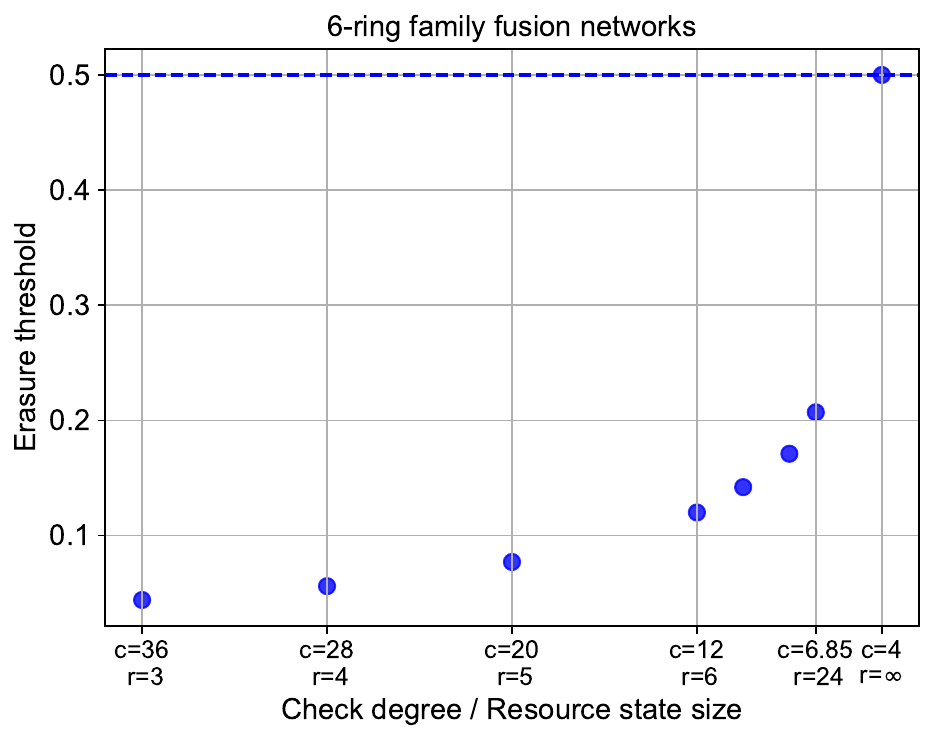} 
	\caption{Resource size vs erasure threshold for diced 6-ring fusion networks. The $x$-axis corresponds to the (largest) number of qubits in each resource state $r$. One can alternatively plot it against the average (un-muxed) check weight $c$. One can interpret each resource state as a brick obtained with a large macromux overhead (so that there are no erasures internal to the brick).  The point $r=6$ corresponds to the six-ring fusion network, and the point $r=24$ corresponds to the six-ring fusion network with the 2x2x2 macromux scheme.  As described in the text, the points with smaller values of $r$ can be derived from the six-ring fusion network by realizing that the six-ring is composed of three 3GHZ and three 3GHX states joined together by fusions.  For example, the point $r=3$ corresponds to treating each of these six intra-six-ring fusions as also being part of the fusion network. The $c=4$ point corresponds to the fusion of unbounded-size toric code resource states.}
	\label{figRSSizeVsThreshold}
\end{figure}

In particular, consider the 6-ring fusion network, with $N$ stage cuboidal dicing. As $N$ increases from 1 to $\infty$, each brick corresponds to resource states of increasing size. In Fig.~\ref{figRSSizeVsThreshold}, we plot the erasure threshold of these schemes for the limit of large macromux overheads $M$, such that there are no erasures in each brick. As $N$ increases, so does the (maximal) resource state size $r$ in terms of the number of qubits. Correspondingly, the average (un-muxed) check degree $c$ decreases. The limit $N\mapsto \infty$ is where resource states are large square-lattice surface codes. In this figure, we also plot the threshold of primitive schemes derived from the 6-ring fusion network by decomposing the resource state to smaller states; $r=3$ corresponds to the fusion network where 6-ring resource states are replaced by six 3-qubit resource states and six fusions between them, $r=4$ corresponds to the fusion network where 6-rings are replaced by three 4-qubit states and three fusions between them, and $r=5$ is where the 6-ring is replaced by a 5-qubit and a 4-qubit state and a two fusions between them.

\begin{figure*}
	\centering
	\includegraphics[width=0.9\columnwidth]{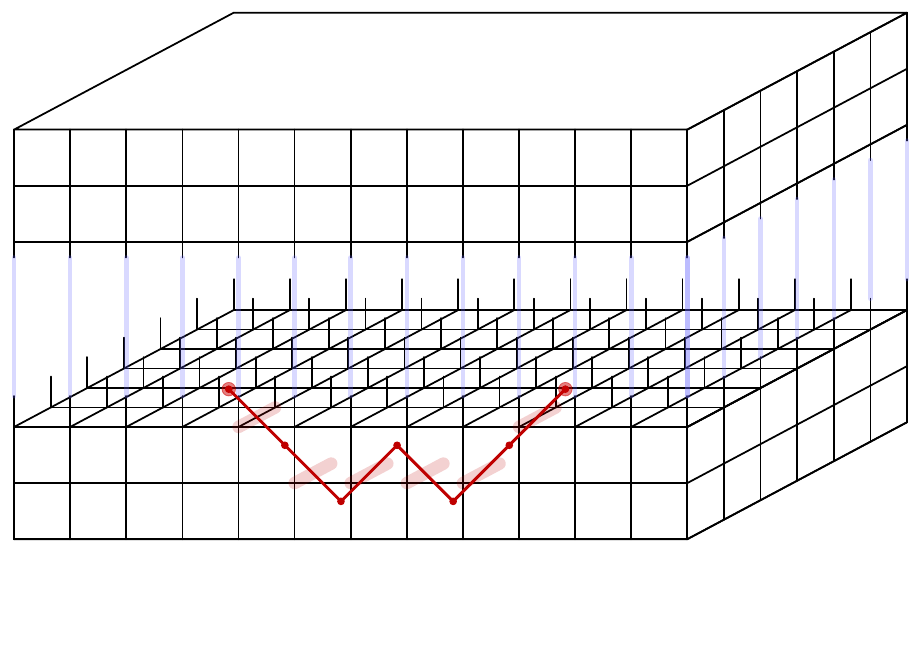}  \qquad
	\includegraphics[width=0.95\columnwidth]{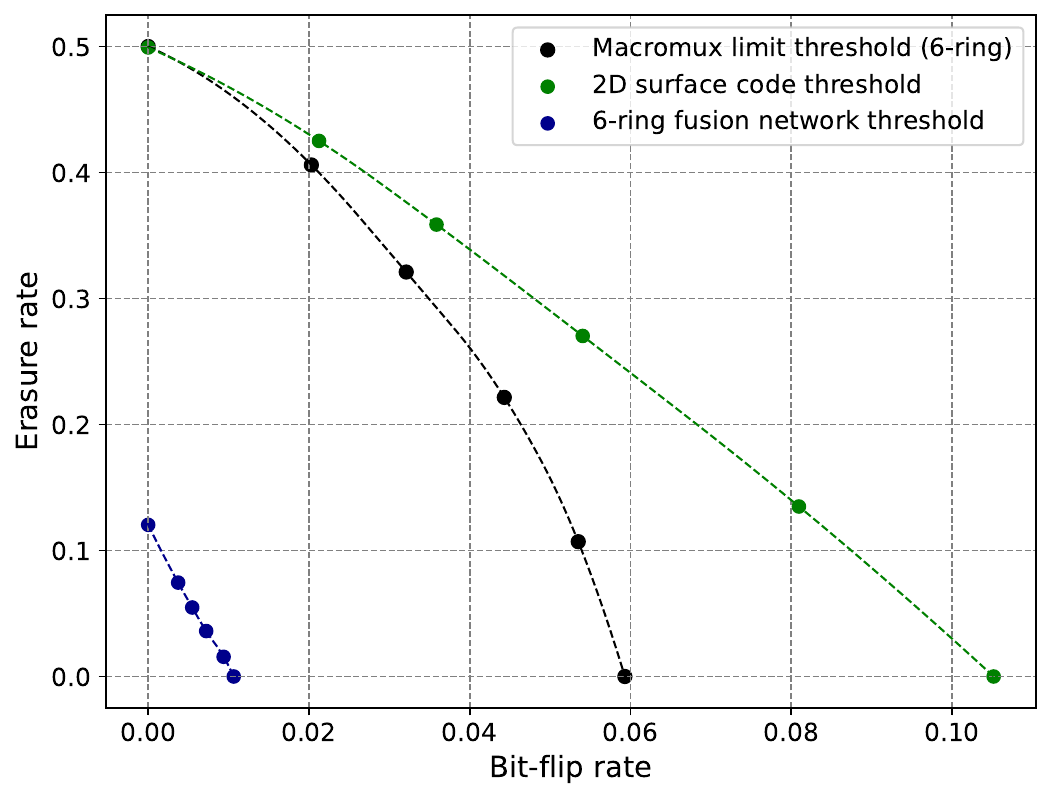} 
	\caption{(left) A pair of large bricks (black cubic lattices) with fusions between them (blue lines). As the brick size gets larger, and the number of copies increases, the only errors that remain (that cannot be postselected away by macromux) are (i) in the interface between bricks, and (ii) error strings that span between these interfaces. As such, the decoding problem approaches that of the 2D surface code (but with a correlated error model).
    (right) Maximum achievable thresholds for the hierarchical macromux scheme with the 6-ring fusion network, as bricks get arbitrarily large and the number of copies tends to infinity. 
    The error model is \textit{i.i.d} bit-flip errors and erasures on all fusion outcomes (a hardware agnotistic circuit-level error model for FBQC). 
    For the 6-ring network, the erasure threshold is $12\%$ and the bit-flip threshold is $1\%$, and the threshold surface approximately interpolates between these marginals~\cite{bartolucci2023fusion}. 
    We also plot the phenomenological threshold of the 2D toric code with perfect measurements under \textit{i.i.d} qubit erasure, bit-flip, and phase-flip (i.e. a code capacity setting), and similarly interpolate a surface as has been studied in Ref.~\cite{stace2010error}. 
    Remarkably, as the brick size and number of copies increases, we approach the $50\%$ erasure threshold of the 2D toric code, and a significant proportion of the 2D surface code bitflip threshold. 
    One can likely improve the threshold by considering degeneracies in the minimal weight corrections ~\cite{stace2009thresholds} when decoding.
    }
	\label{fig2DLimit}
\end{figure*}

Because the boundaries between bricks are two dimensional, the rough intuition is that, as the bricks get bigger, the thresholds should approach the threshold of a two-dimensional surface code with noiseless stabilizer measurements (i.e. the code capacity setting), because macromux lets us choose bricks with no erasures or flipped syndromes. 
However, the behaviour of Pauli and erasure errors are different, as we only receive syndrome information from Pauli errors, and only when there are checks contained within a brick.
So as the bricks become large, the decoding problem does approach that of the 2D surface code, but with a correlated Pauli error model, caused by undetectable spanning errors in the brick. 
Namely, the effective error model contains Pauli errors spanning between any two toric code stabilizers (of the same type) with probability depending on the weight and path multiplicity, along with \textit{i.i.d} erasure errors. 
See Fig.~\ref{fig2DLimit}. 

In Fig.~\ref{fig2DLimit} we show the numerically obtained threshold limits of the scheme. 
The error model considered is \textit{i.i.d.} bit-flip errors and erasures on fusion outcomes of the fusion network. 
This error model may be regarded as a hardware-agnostic circuit-level error model for fusion-based quantum computation. 
We see significant improvements in threshold from the baseline 6-ring fusion network; up to $6\%$ for bit-flip errors and $50\%$ for erasure errors. 
This constitutes an increase of over a factor of $4$ in erasure threshold, and nearly a factor of $6$ in Pauli threshold. 
This is the highest known fault-tolerant threshold that we are aware of, however the resource cost is also unbounded. In the following sections we efficiently realise substantial fractions of these thresholds.

\subsection{Thresholds with finite brick sizes and finite resources}

\begin{figure*}
	\centering
	\includegraphics[width=0.95\columnwidth]{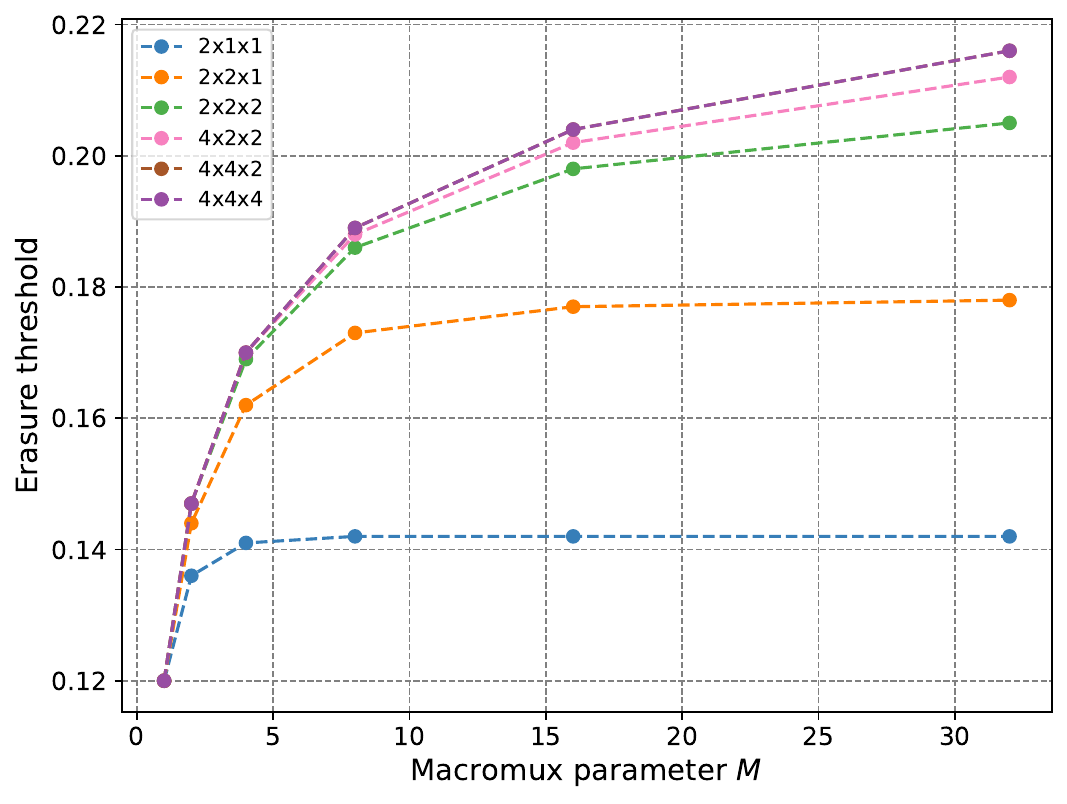} \qquad 
	\includegraphics[width=0.96\columnwidth]{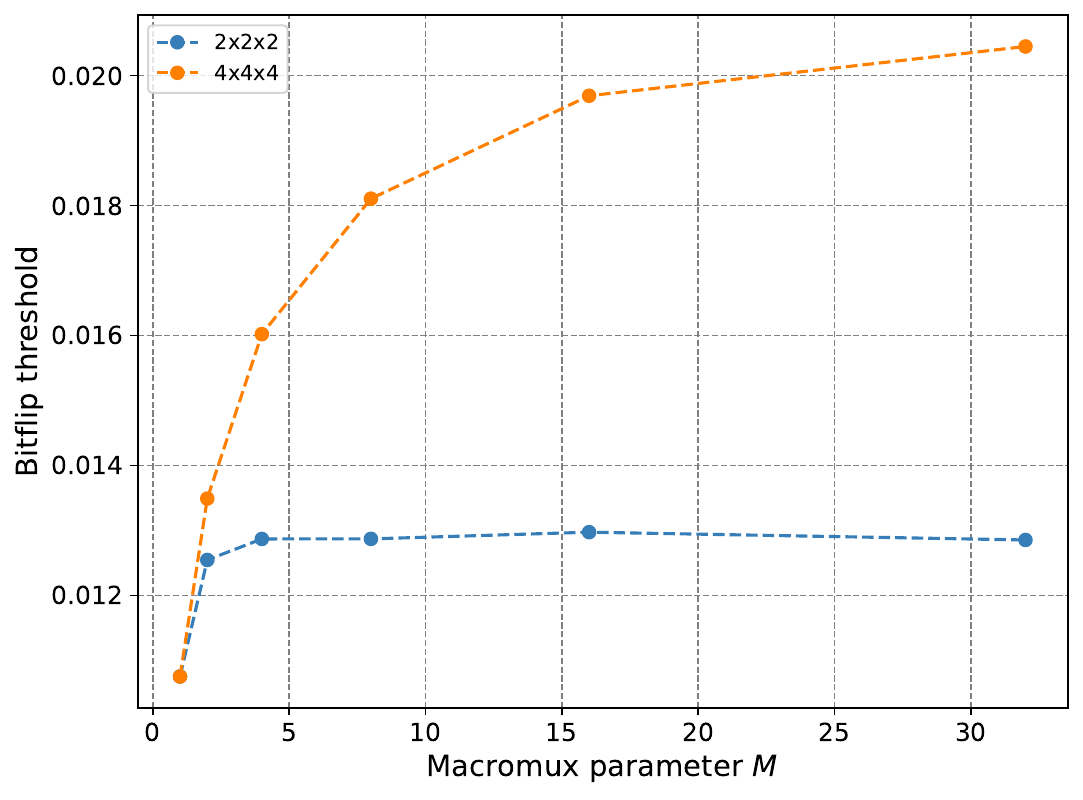} 
	\caption{Thresholds as a function of the macromux parameter $M$ for different maximal brick sizes in the 6-ring fusion network under frozen gap scoring. (left) Erasure-only error model with each outcome subject to erasure with rate $p_E$. (right) Bitflip-only error model with each outcome subject to bitflip with rate $p_P$. Performance with an error model with both erasures and bitflips is given in Fig.~\ref{figmmuxresults2}.}
	\label{figmmuxresults}
\end{figure*}

The previous section considered the limits of macromux as the brick size and number of copies of each brick goes to infinity. 
However, in practice, we are interested in the regime of small numbers of brick copies. 
In the presence of erasures and Pauli errors, the frozen gap scorer allows us to have efficient schemes with high-thresholds. 
In App.~\ref{appK6LO} we show the frozen gap scorer significantly outperforms the count scorer, so we show only results for the former. 
As an example in Fig.~\ref{figmmuxresults} we plot the thresholds of the 3-stage and 6-stage scheme for the six-ring fusion network as a function of number of copies. 
The 3-stage scheme has bricks of size up to 2x2x2 (i.e. 8 resource states), and the 6-stage scheme has bricks of size up to 4x4x4 (i.e. 64 resource states). 

We see that having larger bricks with more stages has significant benefit for the same number of copies.
Furthermore, having larger bricks does not change the overhead required, which only depends on the number of copies. 
We also see that the frozen gap has a significant performance advantage for larger bricks (where more fault-tolerant information and geometry is available), and can nearly double the threshold using a factor of $\sim 10$ additional resource states as compared to no macromux. 

\subsection{Benchmarking high loss-tolerance photonic fusion-based schemes}

Macromux is particularly well suited for photonic fusion-based quantum computing architectures, where it can significantly increase the loss tolerance. Macromux complements other techniques to increase loss tolerance, such as local encoding, local adaptivity and dynamic bias adaptivity~\cite{dynamic_bias, varnava2006loss, azuma2015all, li2015resource, lee2015nearly, ewert2016ultrafastprl, lee2019fundamental, hilaire2021error, bell2023optimizing, lee2023parity, pankovich, songetal}. Several of these schemes have been benchmarked in Ref.~\cite{bartolucci2025comparison}, where the loss tolerance as a function of resource state size (i.e. number of qubits/photons in a resource state) is considered. Importantly, Ref.~\cite{bartolucci2025comparison} points out that resource state size is not a good indicator of the overhead cost to prepare a resource state. For instance, using type-II fusions and 3GHZ states, the overhead cost can scale generally quadratically in the resource state size. On the other hand, macromux requires only linear cost in the macromux overhead $M$.

\begin{figure*}
	\centering
	\includegraphics[width=0.85\columnwidth]{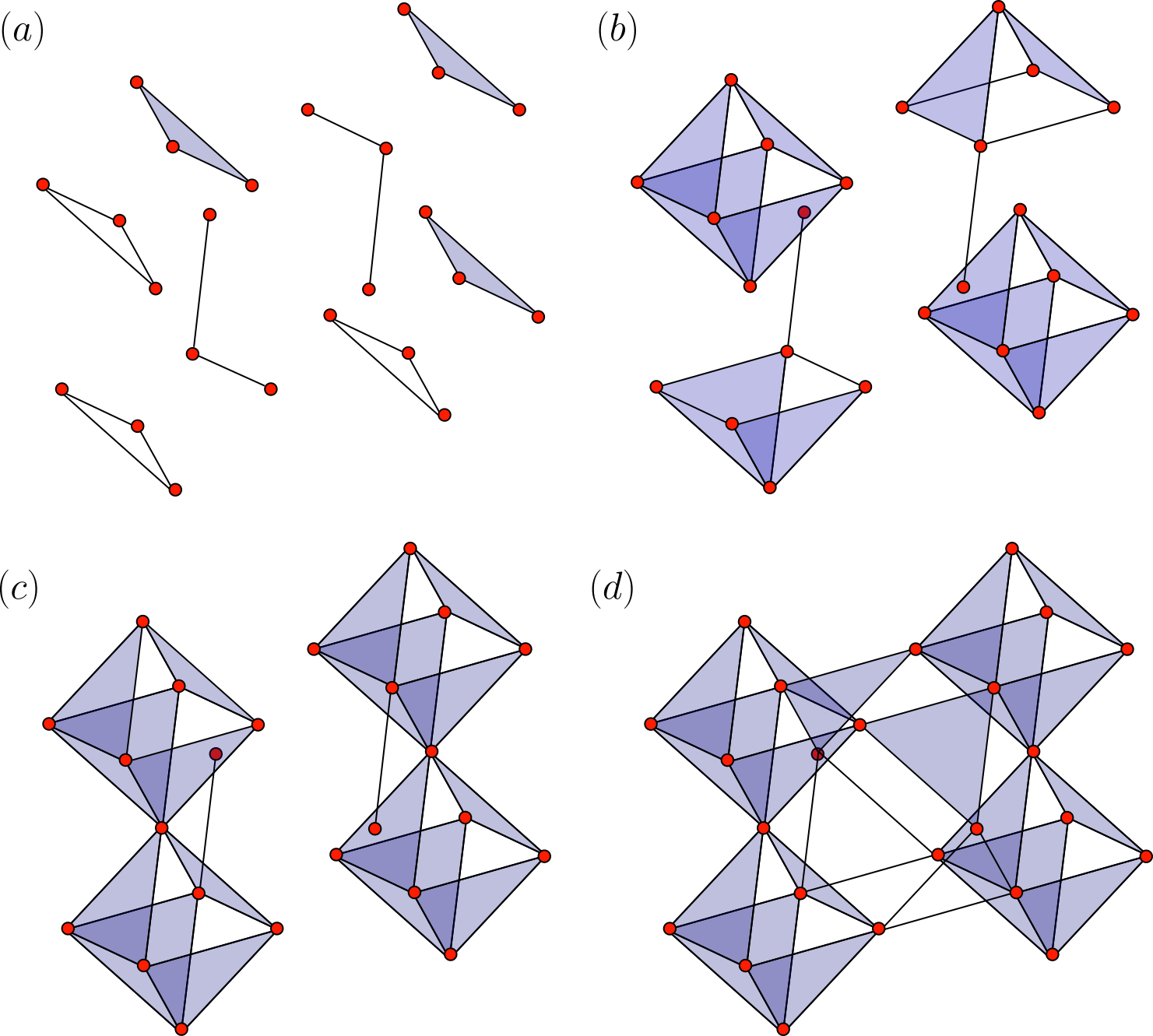} \qquad
	\includegraphics[width=0.95\columnwidth]{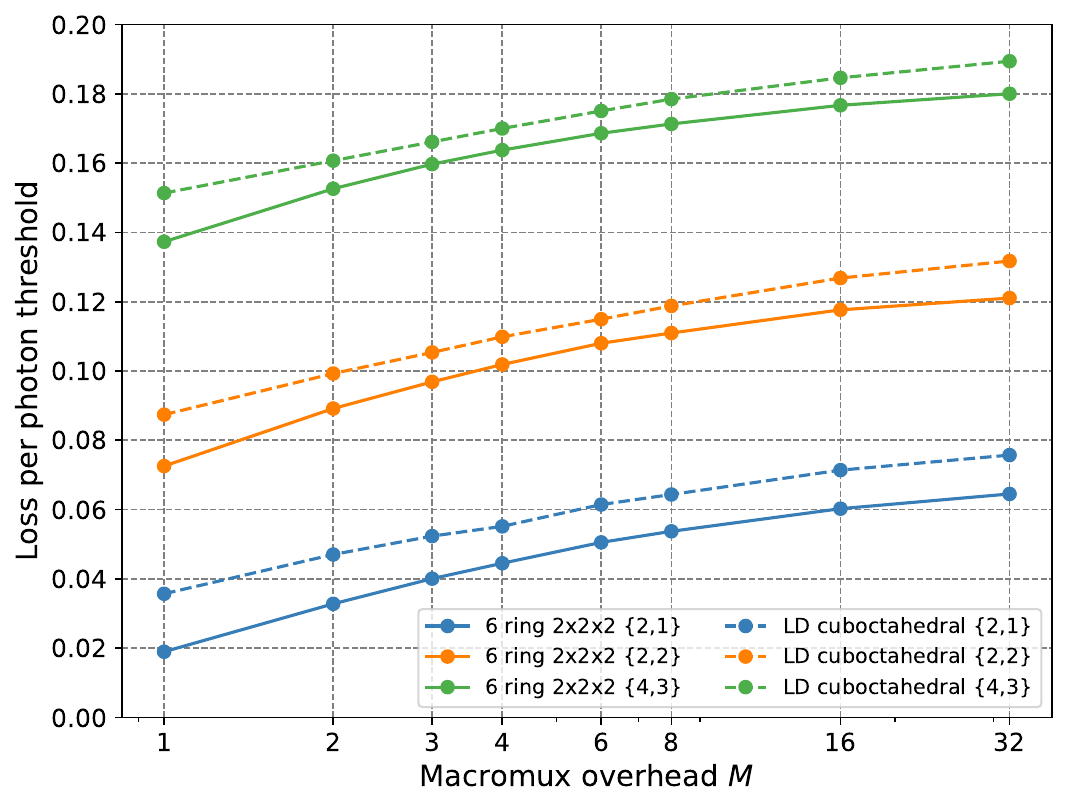} 
	\caption{(left) The (multi-stage) cuboctahedral dicing scheme.  Red vertices are resource states and edges denote fusions between them.  Blue shaded faces are weight-three checks (for more details on the check structure, see \cite{FusionComplexes}).  (a) shows the first stage where groups of three resource states are fused together.  (b) shows the second stage, where each connected component has six resource states fused together.  In the next stage (c) each component has 12 resource states fused together, and finally in (d) we have the final brick comprising 24 resource states and 60 fusions.  For clarity, we have only drawn resource states and fusions involved in a single brick at the fourth (and penultimate) stage of macromux.  The full fusion graph will comprise translates of many of these bricks as well as fusions between them.  This dicing contains a large number of complete checks, which leads to a high tolerance to bit-flip errors with macromux. (right) Loss tolerance thresholds for the 6-ring and loopy diamond resource states with various local encodings, using local adaptivity and dynamic bias arrangement~\cite{dynamic_bias} and the frozen gap scorer. Each fusion is subject to failure, and photon loss (following ~\cite{bartolucci2025comparison}). We consider the 2x2x2 dicing for the 6-ring fusion network, and the cuboctahedral dicing for the loopy diamond fusion network. }
	\label{figbbs}
\end{figure*}

In Fig.~\ref{figbbs}, we show the loss tolerance thresholds for several fusion-based schemes. The error model includes fusion failure and a per-photon loss rate. In addition to the 6-ring hierarchical dicing scheme, we also consider the loopy-diamond fusion network with a cuboctahedral dicing. The loopy diamond fusion network has 8-qubit resource states arranged on a cuboctahedral lattice, as introduced in Ref.~\cite{FusionComplexes} and studied in Ref.~\cite{bartolucci2025comparison}. This dicing scheme is outlined in Fig.~\ref{figbbs}, where checks correspond to the shaded triangles, as well as octahedra and cuboctahedra (for more details, see \cite{FusionComplexes}). While we only present loss tolerance results for this scheme, we note that it provides significant protection from Pauli errors, due to the small, weight 3 checks that are produced in the first stage of macromux, see Fig.~\ref{figmmuxresults2}.  We see that macromux offers significant improvements in loss threshold with low overheads, working favourably with other loss tolerance gadgets. 

To design the best performing scheme for a given overhead, one must carefully allocate resources between local encodings and macromux, noting that local encodings are particularly good at improving loss tolerance, while macromux is particularly powerful at improving Pauli error tolerance.

\section{Discussion}
For practical quantum computation, we need low logical error rates, so finding protocols with high thresholds is a good first step to enabling fault-tolerant quantum computing. 
Here we gave a scheme to efficiently increase the Pauli and erasure thresholds of a given protocol. 
This is efficient in a practical sense: in some cases, the loss tolerance of a scheme can be doubled with an overhead of 3 (i.e., increasing the number of resource states by a factor of 3). 
Previously, (to our knowledge) multiplexing has only been used to increase erasure tolerance (due to, e.g., fusion failure or loss), but here we have given a scheme for leveraging multiplexing to increase the tolerance to Pauli errors. 

We have focussed on topological fusion networks for their simplicity, but the approach can be readily applied to LDPC-based~\cite{chen2025fusion} or concatenation-based~\cite{litinski2025blocklet} schemes as well which offer significant resource reductions due to their high rates. 
These can lead to highly-resource efficient fusion networks that also have high thresholds. 

We have focused on fusion-based quantum computation, but these ideas can also be used in the circuit model of quantum computation provided that a suitable memory and qubit routing are available. 
For linear optical quantum computers these are basic hardware primitives, but these can also be available for some other architectures, such as trapped ions and neutral atoms. 
In the circuit model, to determine suitable dicing schemes, one can consider the ZX network that describes the fault tolerant protocol~\cite{Bombin2024unifyingflavorsof}, and partition it suitably. Furthermore, these schemes can be readily applied to optical schemes based on continuous variable qubit encodings, such as GKP-based qubits~\cite{gottesman2000encoding}. 
It would also be interesting to consider non-stabilizer information in the scoring functions, for use in non-Clifford gates~\cite{bauer2025planar}. 

\section{Acknowledgements}
We thank Hector Bombin, Jake Bulmer, Daniel Litinski, Fernando Pastawski, Terry Rudolph, Chris Sparrow and all our colleagues at PsiQuantum for fruitful discussions.

\clearpage
\newpage

\bibliographystyle{apsrev4-1}
\bibliography{references}

@misc{endote1,
note   = "{LDPC or concatenated schemes may require more advanced hardware capabilities such as long-range couplers~\cite{yoder2025tour} which may lead to higher error rates. Currently simulated LDPC protocols also have lower observed thresholds than topological code approaches~\cite{bravyi2024high}}",
}

@article{yoder2025tour,
  title={Tour de gross: A modular quantum computer based on bivariate bicycle codes},
  author={Yoder, Theodore J and Schoute, Eddie and Rall, Patrick and Pritchett, Emily and Gambetta, Jay M and Cross, Andrew W and Carroll, Malcolm and Beverland, Michael E},
  journal={arXiv preprint arXiv:2506.03094},
  year={2025}
}

@article{bluvstein2024logical,
  title={Logical quantum processor based on reconfigurable atom arrays},
  author={Bluvstein, Dolev and Evered, Simon J and Geim, Alexandra A and Li, Sophie H and Zhou, Hengyun and Manovitz, Tom and Ebadi, Sepehr and Cain, Madelyn and Kalinowski, Marcin and Hangleiter, Dominik and others},
  journal={Nature},
  volume={626},
  number={7997},
  pages={58--65},
  year={2024},
  publisher={Nature Publishing Group UK London}
}

@article{google2025quantum,
  title={Quantum error correction below the surface code threshold},
  journal={Nature},
  volume={638},
  number={8052},
  pages={920--926},
  year={2025},
  publisher={Nature Publishing Group UK London}
}

@article{paetznick2024demonstration,
  title={Demonstration of logical qubits and repeated error correction with better-than-physical error rates},
  author={Paetznick, A and Da Silva, MP and Ryan-Anderson, C and Bello-Rivas, JM and Campora III, JP and Chernoguzov, A and Dreiling, JM and Foltz, C and Frachon, F and Gaebler, JP and others},
  journal={arXiv preprint arXiv:2404.02280},
  year={2024}
}

@article{reichardt2024demonstration,
  title={Demonstration of quantum computation and error correction with a tesseract code},
  author={Reichardt, Ben W and Aasen, David and Chao, Rui and Chernoguzov, Alex and van Dam, Wim and Gaebler, John P and Gresh, Dan and Lucchetti, Dominic and Mills, Michael and Moses, Steven A and others},
  journal={arXiv preprint arXiv:2409.04628},
  year={2024}
}

@article{larsen2025integrated,
  title={Integrated photonic source of Gottesman--Kitaev--Preskill qubits},
  author={Larsen, Mikkel V and Bourassa, J Eli and Kocsis, Sacha and Tasker, Joel F and Chadwick, Robert S and Gonz{\'a}lez-Arciniegas, Carlos and Hastrup, Jacob and Lopetegui-Gonz{\'a}lez, Carlos E and Miatto, Filippo M and Motamedi, A and others},
  journal={Nature},
  volume={642},
  number={8068},
  pages={587--591},
  year={2025},
  publisher={Nature Publishing Group UK London}
}

@article{roffe_decoding_2020,
   title={Decoding across the quantum low-density parity-check code landscape},
   volume={2},
   ISSN={2643-1564},
   url={http://dx.doi.org/10.1103/PhysRevResearch.2.043423},
   DOI={10.1103/physrevresearch.2.043423},
   number={4},
   journal={Physical Review Research},
   publisher={American Physical Society (APS)},
   author={Roffe, Joschka and White, David R. and Burton, Simon and Campbell, Earl},
   year={2020},
   month={Dec}
}

@article{lacroix2025scaling,
  title={Scaling and logic in the colour code on a superconducting quantum processor},
  author={Lacroix, Nathan and Bourassa, Alexandre and Heras, Francisco JH and Zhang, Lei M and Bausch, Johannes and Senior, Andrew W and Edlich, Thomas and Shutty, Noah and Sivak, Volodymyr and Bengtsson, Andreas and others},
  journal={Nature},
  volume={645},
  number={8081},
  pages={614--619},
  year={2025},
  publisher={Nature Publishing Group UK London}
}

@article{hutter2014efficient,
  title={Efficient Markov chain Monte Carlo algorithm for the surface code},
  author={Hutter, Adrian and Wootton, James R and Loss, Daniel},
  journal={Physical Review A},
  volume={89},
  number={2},
  pages={022326},
  year={2014},
  publisher={APS}
}

@article{songetal,
  title = {Encoded-Fusion-Based Quantum Computation for High Thresholds with Linear Optics},
  author = {Song, Wooyeong and Kang, Nuri and Kim, Yong-Su and Lee, Seung-Woo},
  journal = {Phys. Rev. Lett.},
  volume = {133},
  issue = {5},
  pages = {050605},
  numpages = {6},
  year = {2024},
  month = {Aug},
  publisher = {American Physical Society},
  doi = {10.1103/PhysRevLett.133.050605},
  url = {https://link.aps.org/doi/10.1103/PhysRevLett.133.050605},
  note={arXiv:2408.01041 [quant-ph]}
}

@article{pankovich,
  title={High-Photon-Loss Threshold Quantum Computing Using GHZ-State Measurements},
  author={Pankovich, Brendan and Kan, Angus and Wan, Kwok Ho and Ostmann, Maike and Neville, Alex and Omkar, Srikrishna and Sohbi, Adel and Br{\'a}dler, Kamil},
  journal={Physical Review Letters},
  volume={133},
  number={5},
  pages={050604},
  year={2024},
  publisher={APS}
}

@article{hilaire2021error,
  title={Error-correcting entanglement swapping using a practical logical photon encoding},
  author={Hilaire, Paul and Barnes, Edwin and Economou, Sophia E and Grosshans, Fr{\'e}d{\'e}ric},
  journal={Physical Review A},
  volume={104},
  number={5},
  pages={052623},
  year={2021},
  publisher={APS}
}

@article{lee2019fundamental,
  title={Fundamental building block for all-optical scalable quantum networks},
  author={Lee, Seung-Woo and Ralph, Timothy C and Jeong, Hyunseok},
  journal={Physical Review A},
  volume={100},
  number={5},
  pages={052303},
  year={2019},
  publisher={APS}
}

@article{ewert2016ultrafastprl,
  title={Ultrafast long-distance quantum communication with static linear optics},
  author={Ewert, Fabian and Bergmann, Marcel and van Loock, Peter},
  journal={Physical review letters},
  volume={117},
  number={21},
  pages={210501},
  year={2016},
  publisher={APS}
}

@article{lee2023parity,
  title={Parity-encoding-based quantum computing with Bayesian error tracking},
  author={Lee, Seok-Hyung and Omkar, Srikrishna and Teo, Yong Siah and Jeong, Hyunseok},
  journal={npj Quantum Information},
  volume={9},
  number={1},
  pages={39},
  year={2023},
  publisher={Nature Publishing Group UK London}
}

@article{lee2015nearly,
  title={Nearly deterministic Bell measurement for multiphoton qubits and its application to quantum information processing},
  author={Lee, Seung-Woo and Park, Kimin and Ralph, Timothy C and Jeong, Hyunseok},
  journal={Physical review letters},
  volume={114},
  number={11},
  pages={113603},
  year={2015},
  publisher={APS}
}

@article{li2015resource,
  title={Resource costs for fault-tolerant linear optical quantum computing},
  author={Li, Ying and Humphreys, Peter C. and Mendoza, Gabriel J. and Benjamin, Simon C.},
  journal={Physical Review X},
  volume={5},
  number={4},
  pages={041007},
  year={2015},
  publisher={APS}
}

@article{azuma2015all,
  title={All-photonic quantum repeaters},
  author={Azuma, Koji and Tamaki, Kiyoshi and Lo, Hoi-Kwong},
  journal={Nature communications},
  volume={6},
  number={1},
  pages={1--7},
  year={2015},
  publisher={Nature Publishing Group}
}

@article{varnava2006loss,
  title={Loss tolerance in one-way quantum computation via counterfactual error correction},
  author={Varnava, Michael and Browne, Daniel E and Rudolph, Terry},
  journal={Physical review letters},
  volume={97},
  number={12},
  pages={120501},
  year={2006},
  publisher={APS}
}

@article{bell2023optimizing,
  title={Optimizing graph codes for measurement-based loss tolerance},
  author={Bell, Thomas J and Pettersson, Love A and Paesani, Stefano},
  journal={PRX Quantum},
  volume={4},
  number={2},
  pages={020328},
  year={2023},
  publisher={APS}
}

@article{yoshida2025concatenate,
  title={Concatenate codes, save qubits},
  author={Yoshida, Satoshi and Tamiya, Shiro and Yamasaki, Hayata},
  journal={npj Quantum Information},
  volume={11},
  number={1},
  pages={88},
  year={2025},
  publisher={Nature Publishing Group UK London}
}

@article{aliferis2005quantum,
  title={Quantum accuracy threshold for concatenated distance-3 codes},
  author={Aliferis, Panos and Gottesman, Daniel and Preskill, John},
  journal={arXiv preprint quant-ph/0504218},
  year={2005}
}

@misc{dynamic_bias,
  author = {Hector Bombín and Chris Dawson and Naomi Nickerson and Mihir Pant and Jordan Sullivan},
  title = {{I}ncreasing error tolerance in quantum computers with dynamic bias arrangement},
  year = {2023},
  eprint = {2303.16122},
  note = {arXiv:2303.16122v1}
}

@inproceedings{aharonov1997fault,
  title={Fault-tolerant quantum computation with constant error},
  author={Aharonov, Dorit and Ben-Or, Michael},
  booktitle={Proceedings of the twenty-ninth annual ACM symposium on Theory of computing},
  pages={176--188},
  year={1997}
}

@article{knill1996concatenated,
  title={Concatenated quantum codes},
  author={Knill, Emanuel and Laflamme, Raymond},
  journal={arXiv preprint quant-ph/9608012},
  year={1996}
}

@article{bartolucci2025comparison,
  title={Comparison of schemes for highly loss tolerant photonic fusion based quantum computing},
  author={Bartolucci, Sara and Bell, Tom and Bombin, Hector and Birchall, Patrick and Bulmer, Jacob and Dawson, Christopher and Farrelly, Terry and Gartenstein, Samuel and Gimeno-Segovia, Mercedes and Litinski, Daniel and others},
  journal={arXiv preprint arXiv:2506.11975},
  year={2025}
}

@article{stace2010error,
  title={Error correction and degeneracy in surface codes suffering loss},
  author={Stace, Thomas M and Barrett, Sean D},
  journal={Physical Review A—Atomic, Molecular, and Optical Physics},
  volume={81},
  number={2},
  pages={022317},
  year={2010},
  publisher={APS}
}

@article{jacob2025single,
  title={Single-shot decoding and fault-tolerant gates with trivariate tricycle codes},
  author={Jacob, Abraham and McLauchlan, Campbell and Browne, Dan E},
  journal={arXiv preprint arXiv:2508.08191},
  year={2025}
}

@article{xie2026simple,
  title={Simple, Efficient, and Generic Post-Selection Decoding for qLDPC codes},
  author={Xie, Haipeng and Yoshioka, Nobuyuki and Tsubouchi, Kento and Li, Ying},
  journal={arXiv preprint arXiv:2601.17757},
  year={2026}
}

@article{bartolucci2021switch,
  title={Switch networks for photonic fusion-based quantum computing},
  author={Bartolucci, Sara and Birchall, Patrick and Bonneau, Damien and Cable, Hugo and Gimeno-Segovia, Mercedes and Kieling, Konrad and Nickerson, Naomi and Rudolph, Terry and Sparrow, Chris},
  journal={arXiv preprint arXiv:2109.13760},
  year={2021}
}

@article{chen2025fusion,
  title={Fusion-based implementation of qLDPC codes with quantum emitters},
  author={Chen, Susan X and L{\"o}bl, Matthias C and Chan, Ming Lai and S{\o}rensen, Anders S and Paesani, Stefano},
  journal={arXiv preprint arXiv:2509.17223},
  year={2025}
}

@article{bauer2025planar,
  title={Planar fault-tolerant circuits for non-Clifford gates on the 2D color code},
  author={Bauer, Andreas and de la Fuente, Julio C Magdalena},
  journal={arXiv preprint arXiv:2505.05175},
  year={2025}
}

@article{gottesman2000encoding,
  title={Encoding a qubit in an oscillator},
  author={Gottesman, Daniel and Kitaev, Alexei and Preskill, John},
  journal={arXiv preprint quant-ph/0008040},
  year={2000}
}

@article{bravyi2024high,
  title={High-threshold and low-overhead fault-tolerant quantum memory},
  author={Bravyi, Sergey and Cross, Andrew W and Gambetta, Jay M and Maslov, Dmitri and Rall, Patrick and Yoder, Theodore J},
  journal={Nature},
  volume={627},
  number={8005},
  pages={778--782},
  year={2024},
  publisher={Nature Publishing Group UK London}
}

@inproceedings{panteleev2022asymptotically,
  title={Asymptotically good quantum and locally testable classical LDPC codes},
  author={Panteleev, Pavel and Kalachev, Gleb},
  booktitle={Proceedings of the 54th annual ACM SIGACT symposium on theory of computing},
  pages={375--388},
  year={2022}
}

@article{panteleev2021degenerate,
  title={Degenerate quantum LDPC codes with good finite length performance},
  author={Panteleev, Pavel and Kalachev, Gleb},
  journal={Quantum},
  volume={5},
  pages={585},
  year={2021},
  publisher={Verein zur F{\"o}rderung des Open Access Publizierens in den Quantenwissenschaften}
}

@article{litinski2025blocklet,
  title={Blocklet concatenation: Low-overhead fault-tolerant protocols for fusion-based quantum computation},
  author={Litinski, Daniel},
  journal={arXiv preprint arXiv:2506.13619},
  year={2025}
}

@article{yamasaki2024time,
  title={Time-efficient constant-space-overhead fault-tolerant quantum computation},
  author={Yamasaki, Hayata and Koashi, Masato},
  journal={Nature Physics},
  volume={20},
  number={2},
  pages={247--253},
  year={2024},
  publisher={Nature Publishing Group UK London}
}

@article{tamiya2024polylog,
  title={Polylog-time-and constant-space-overhead fault-tolerant quantum computation with quantum low-density parity-check codes},
  author={Tamiya, Shiro and Koashi, Masato and Yamasaki, Hayata},
  journal={arXiv preprint arXiv:2411.03683},
  year={2024}
}

@article{akahoshi2025runtime,
  title={Runtime reduction in lattice surgery utilizing time-like soft information},
  author={Akahoshi, Yutaro and Toshio, Riki and Fujisaki, Jun and Oshima, Hirotaka and Sato, Shintaro and Fujii, Keisuke},
  journal={arXiv preprint arXiv:2510.21149},
  year={2025}
}

@article{lee2025efficient,
  title={Efficient Post-Selection for General Quantum LDPC Codes},
  author={Lee, Seok-Hyung and English, Lucas and Bartlett, Stephen D},
  journal={arXiv preprint arXiv:2510.05795},
  year={2025}
}

@article{chen2025scalable,
  title={Scalable accuracy gains from postselection in quantum error correcting codes},
  author={Chen, Hongkun and Xu, Daohong and Sommers, Grace M and Huse, David A and Thompson, Jeff D and Gopalakrishnan, Sarang},
  journal={arXiv preprint arXiv:2510.05222},
  year={2025}
}

@article{PsiHardware,
  title={A manufacturable platform for photonic quantum computing},
  author={{PsiQuantum team}},
  journal={Nature},
  volume={641},
  pages={876–883},
  year={2025},
}

@article{FusionComplexes,
  title={Fault-tolerant complexes},
  author={Bombin, Hector and Dawson, Chris and Farrelly, Terry and Liu, Yehua and Nickerson, Naomi and Pant, Mihir and Pastawski, Fernando and Roberts, Sam},
  journal={arXiv preprint arXiv:2308.07844},
  year={2023}
}

@article{pattison2023hierarchical,
  title={Hierarchical memories: Simulating quantum ldpc codes with local gates},
  author={Pattison, Christopher A and Krishna, Anirudh and Preskill, John},
  journal={arXiv preprint arXiv:2303.04798},
  year={2023}
}

@article{Bombin2024unifyingflavorsof,
  doi = {10.22331/q-2024-06-18-1379},
  url = {https://doi.org/10.22331/q-2024-06-18-1379},
  title = {Unifying flavors of fault tolerance with the {ZX} calculus},
  author = {Bombin, Hector and Litinski, Daniel and Nickerson, Naomi and Pastawski, Fernando and Roberts, Sam},
  journal = {{Quantum}},
  issn = {2521-327X},
  publisher = {{Verein zur F{\"{o}}rderung des Open Access Publizierens in den Quantenwissenschaften}},
  volume = {8},
  pages = {1379},
  year = {2024}
}

@article{gidney2023yoked,
  title={Yoked surface codes},
  author={Gidney, Craig and Newman, Michael and Brooks, Peter and Jones, Cody},
  journal={arXiv preprint arXiv:2312.04522},
  year={2023}
}

@article{bombin2023logical,
  title={Logical blocks for fault-tolerant topological quantum computation},
  author={Bombin, Hector and Dawson, Chris and Mishmash, Ryan V and Nickerson, Naomi and Pastawski, Fernando and Roberts, Sam},
  journal={PRX Quantum},
  volume={4},
  number={2},
  pages={020303},
  year={2023},
  publisher={APS}
}

@article{breuckmann2021balanced,
  title={Balanced product quantum codes},
  author={Breuckmann, Nikolas P and Eberhardt, Jens N},
  journal={IEEE Transactions on Information Theory},
  volume={67},
  number={10},
  pages={6653--6674},
  year={2021},
  publisher={IEEE}
}

@article{panteleev2021asymptotically,
  title={Asymptotically Good Quantum and Locally Testable Classical LDPC Codes},
  author={Panteleev, Pavel and Kalachev, Gleb},
  journal={arXiv preprint arXiv:2111.03654},
  year={2021}
}

@article{landahl2011fault,
  title={Fault-tolerant quantum computing with color codes},
  author={Landahl, Andrew J and Anderson, Jonas T and Rice, Patrick R},
  journal={arXiv preprint arXiv:1108.5738},
  url={https://arxiv.org/abs/1108.5738},
  year={2011}
}

@inproceedings{coecke2008interacting,
  title={Interacting quantum observables},
  author={Coecke, Bob and Duncan, Ross},
  booktitle={International Colloquium on Automata, Languages, and Programming},
  pages={298--310},
  year={2008},
  organization={Springer}
}

@article{van2020zx,
  title={ZX-calculus for the working quantum computer scientist},
  author={van de Wetering, John},
  journal={arXiv preprint arXiv:2012.13966},
  year={2020}
}

@article{stace2009thresholds,
  title={Thresholds for topological codes in the presence of loss},
  author={Stace, Thomas M and Barrett, Sean D and Doherty, Andrew C},
  journal={Physical review letters},
  volume={102},
  number={20},
  pages={200501},
  year={2009},
  publisher={APS}
}

@article{bombin2006topological,
  title={Topological quantum distillation},
  author={Bombin, Hector and Martin-Delgado, Miguel Angel},
  journal={Physical review letters},
  volume={97},
  number={18},
  pages={180501},
  year={2006},
  publisher={APS}
}

@article{tillich2014quantum,
  title={Quantum LDPC codes with positive rate and minimum distance proportional to the square root of the blocklength},
  author={Tillich, Jean-Pierre and Z{\'e}mor, Gilles},
  journal={IEEE Transactions on Information Theory},
  volume={60},
  number={2},
  pages={1193--1202},
  year={2014},
  publisher={IEEE}
}

@article{breuckmann2021ldpc,
  title = {Quantum Low-Density Parity-Check Codes},
  author = {Breuckmann, Nikolas P. and Eberhardt, Jens Niklas},
  journal = {PRX Quantum},
  volume = {2},
  issue = {4},
  pages = {040101},
  numpages = {19},
  year = {2021},
  month = {Oct},
  publisher = {American Physical Society},
  doi = {10.1103/PRXQuantum.2.040101},
  url = {https://link.aps.org/doi/10.1103/PRXQuantum.2.040101}
}

@incollection{kitaev1997quantum,
	title={Quantum error correction with imperfect gates},
	author={Kitaev, A Yu},
	booktitle={Quantum Communication, Computing, and Measurement},
	pages={181--188},
	year={1997},
	publisher={Springer}
}

@article{dennis2002topological,
	title={Topological quantum memory},
	author={Dennis, Eric and Kitaev, Alexei and Landahl, Andrew and Preskill, John},
	journal={Journal of Mathematical Physics},
	volume={43},
	number={9},
	pages={4452--4505},
	year={2002},
	publisher={American Institute of Physics}
}

@article{bombin2021interleaving,
  title={Interleaving: Modular architectures for fault-tolerant photonic quantum computing},
  author={Bombin, Hector and Kim, Isaac H and Litinski, Daniel and Nickerson, Naomi and Pant, Mihir and Pastawski, Fernando and Roberts, Sam and Rudolph, Terry},
  journal={arXiv preprint arXiv:2103.08612},
  url={https://arxiv.org/abs/2103.08612},
  year={2021}
}

@article{bartolucci2023fusion,
  title={Fusion-based quantum computation},
  author={Bartolucci, Sara and Birchall, Patrick and Bombin, Hector and Cable, Hugo and Dawson, Chris and Gimeno-Segovia, Mercedes and Johnston, Eric and Kieling, Konrad and Nickerson, Naomi and Pant, Mihir and others},
  journal={Nature Communications},
  volume={14},
  number={1},
  pages={912},
  year={2023},
  publisher={Nature Publishing Group UK London}
}

@article{BR04,
  title = {Resource-Efficient Linear Optical Quantum Computation},
  author = {Browne, Daniel E. and Rudolph, Terry},
  journal = {Phys. Rev. Lett.},
  volume = {95},
  issue = {1},
  pages = {010501},
  numpages = {4},
  year = {2005},
  month = {Jun},
  publisher = {American Physical Society},
  doi = {10.1103/PhysRevLett.95.010501},
  url = {https://link.aps.org/doi/10.1103/PhysRevLett.95.010501},
  eprint = {quant-ph/0405157},
  archivePrefix = {arXiv},
}

@article{meister2024efficient,
  title={Efficient soft-output decoders for the surface code},
  author={Meister, Nadine and Pattison, Christopher A and Preskill, John},
  journal={arXiv preprint arXiv:2405.07433},
  year={2024}
}

@article{smith2024mitigating,
  title={Mitigating errors in logical qubits},
  author={Smith, Samuel C and Brown, Benjamin J and Bartlett, Stephen D},
  journal={Communications Physics},
  volume={7},
  number={1},
  pages={386},
  year={2024},
  publisher={Nature Publishing Group UK London}
}

@article{bombin2024fault,
  title={Fault-tolerant postselection for low-overhead magic state preparation},
  author={Bomb{\'\i}n, H{\'e}ctor and Pant, Mihir and Roberts, Sam and Seetharam, Karthik I},
  journal={PRX Quantum},
  volume={5},
  number={1},
  pages={010302},
  year={2024},
  publisher={APS}
}

@article{english2024thresholds,
  title={Thresholds for post-selected quantum error correction from statistical mechanics},
  author={English, Lucas H and Williamson, Dominic J and Bartlett, Stephen D},
  journal={arXiv preprint arXiv:2410.07598},
  year={2024}
}

@article{gidney2024magic,
  title={Magic state cultivation: growing T states as cheap as CNOT gates},
  author={Gidney, Craig and Shutty, Noah and Jones, Cody},
  journal={arXiv preprint arXiv:2409.17595},
  year={2024}
}

@article{Knill05,
  title={Quantum computing with realistically noisy devices},
  author={Knill, Emanuel},
  journal={Nature},
  volume={434},
  pages={39-44},
  year={2005},
  publisher={Nature Publishing Group}
}

\clearpage
\onecolumngrid
\appendix

\section{Evaluating logical and frozen gaps}\label{appFG}

\begin{figure}[b]
	\centering
	\includegraphics[width=0.45\columnwidth]{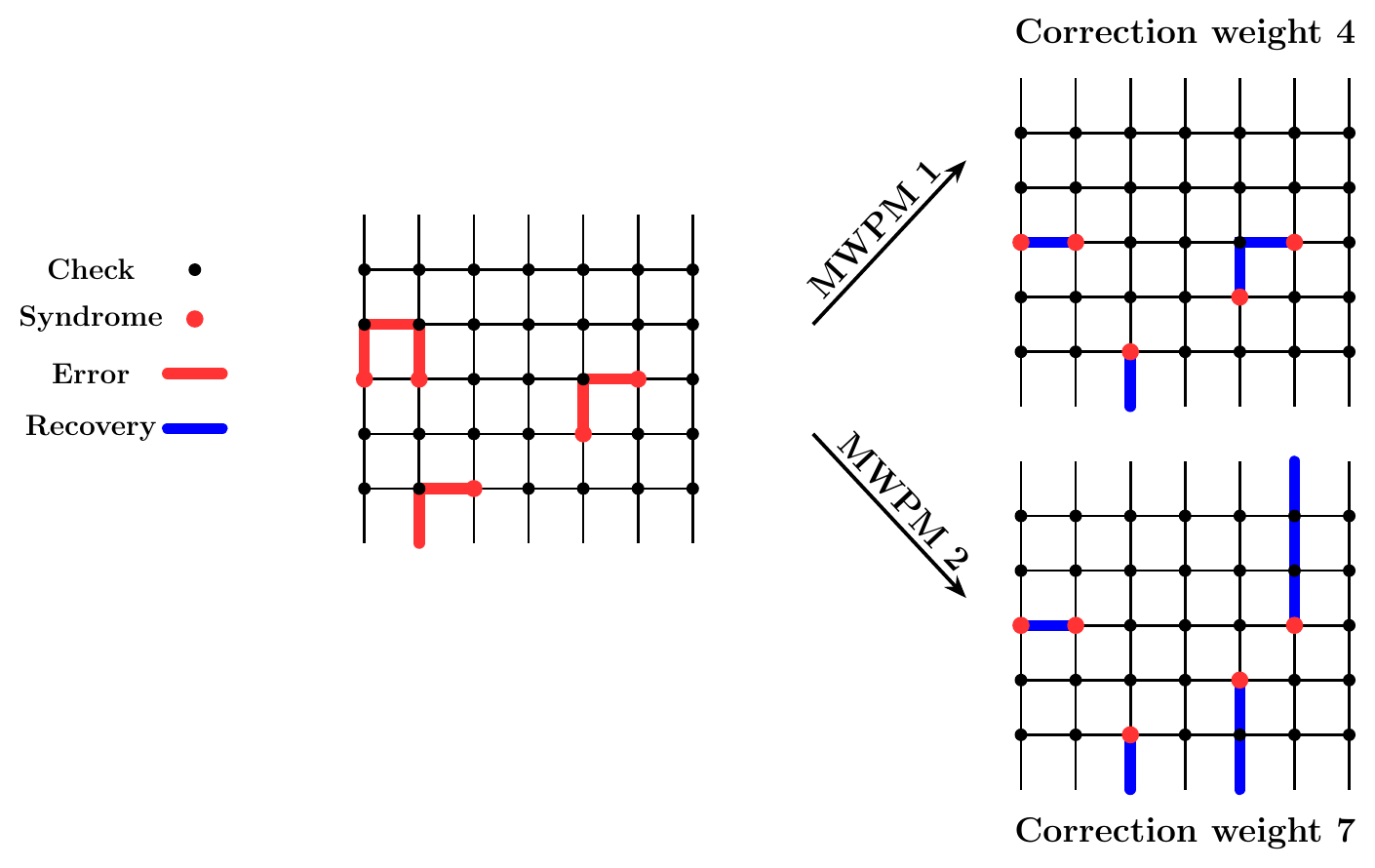} \qquad 
	\includegraphics[width=0.4\columnwidth]{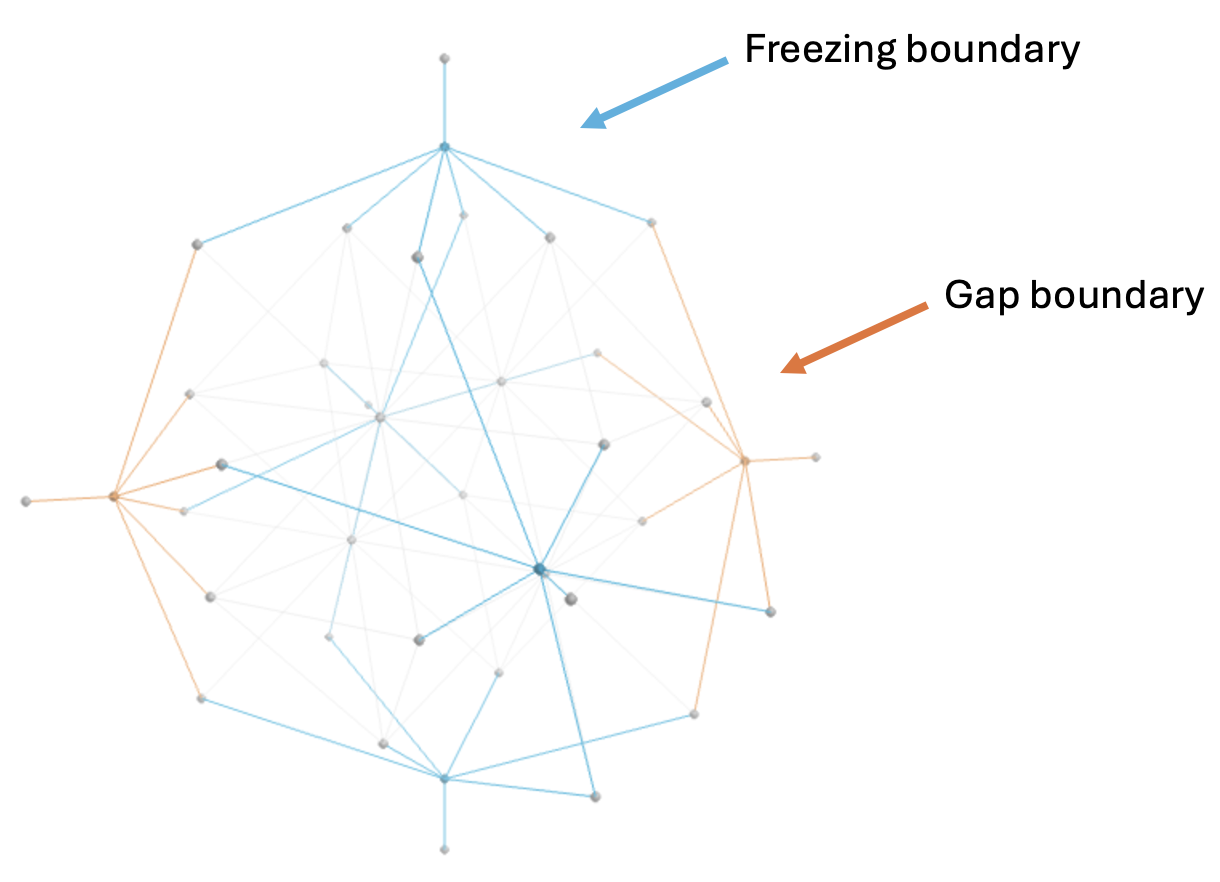} 
	\caption{(left) Example of computing the logical gap on a 2D surface code. MWPM1 and MWPM2 produce the two minimum-weight recoveries in each logical sector.  (The corrections being in different logical sectors means that both corrections remove the syndrome, but they have different effects on the logical codespace.) The logical gap here is $\Delta = |4-8| = 4$. (right) Example of the 3D syndrome graph for a 3x3x3 brick, with freezing and gap boundaries.}
	\label{figGapCalculation2}
\end{figure}

\textbf{Logical gap calculation.}
In Fig.~\ref{figGapCalculation2} we show an example of the gap calculation on a 2D syndrome graph. When all checks have been measured (i.e., there is complete information), there are well-defined logicals and logical sectors from which one can compute two minimal weight corrections. 

\textbf{Logical gap calculation for general protocols.}
Calculating the logical gap for surface-code protocols with boundaries is straightforward, as we can leverage the syndrome graph and MWPM method in the previous section. For general codes, we can modify the calculation using an augmented Tanner graph.

Given a fault-tolerant protocol or quantum error correcting code with a Tanner graph $G$, consisting of check nodes and outcome nodes, let $L_1$, $L_2$, ... $L_k$ be a set of logical membrane representatives. Logical membranes are the objects dual to logical errors; for codes they are logical operators, while for protocols they can be thought of as the logical operator in space-time (see for example Ref.~\cite{bombin2023logical} for more details). We construct the augmented Tanner graph $G_i'$ for each logical $L_i$ as follows:
\begin{itemize}
    \item Add a pseudo check node $v_i$ to the Tanner graph. \item $v_i$ is connected by an edge to each outcome node in the support of $L_i$.
\end{itemize}

Given the augmented Tanner graph $G_i'$ and a syndrome and erasure configuration, we can calculate a minimal weight correction in each logical equivalence class, by setting the pseudo-checks $v_i \mapsto 0,1$, and decoding the Tanner graph. We denote $w_{i,0}$ and $w_{i,1}$ as the minimum weight corrections for decoding on the augmented graph $G_i'$ with pseudo check nodes set to $0$ and $1$, respectively. The logical gap for the $i$th logical equivalence class is given by $\Delta_i = |w_{i,0} - w_{i,1}|$. These gaps can be used to calculate a score, for example  
\begin{equation}
    S = -\sum_{i=1}^k e^{-\alpha \Delta_i},
\end{equation}
but many variants are possible.

Note that in general, computing minimal-weight corrections is computationally expensive, and often non-efficient integer-linear-programming (ILP) methods are deployed~\cite{landahl2011fault,lacroix2025scaling}. As an alternative heuristic approach to calculating the gap, one can in practice use other faster decoders, such as BP-OSD~\cite{roffe_decoding_2020} which do not necessarily return a minimal weight recovery.

We remark that rather than looking at each sector independently, one may look at the minimal weight recoveries for all $2^k$ inequivalent logical sectors. This may be particularly important for higher rate codes where the correlated logical errors may be important to consider. In particular, one may construct the global augmented Tanner graph $G'$ as follows.
\begin{itemize}
    \item For each logical membrane $L_i \in \{L_1, L_2, \ldots, L_k$\}, add a pseudo check node $v_i$ to the Tanner graph. 
    \item Each node $v_i$ is connected by an edge to each outcome node in the support of $L_i$.
\end{itemize}

\begin{figure}
	\centering
	\includegraphics[width=0.425\columnwidth]{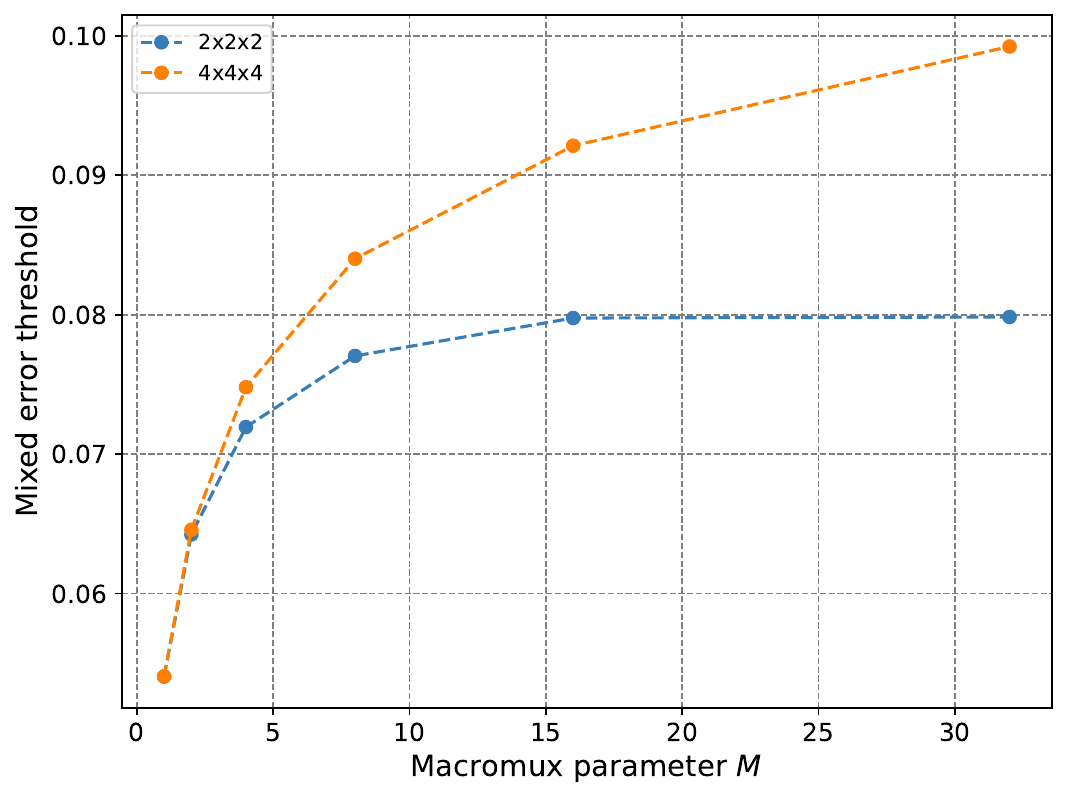} \qquad 
	\includegraphics[width=0.425\columnwidth]{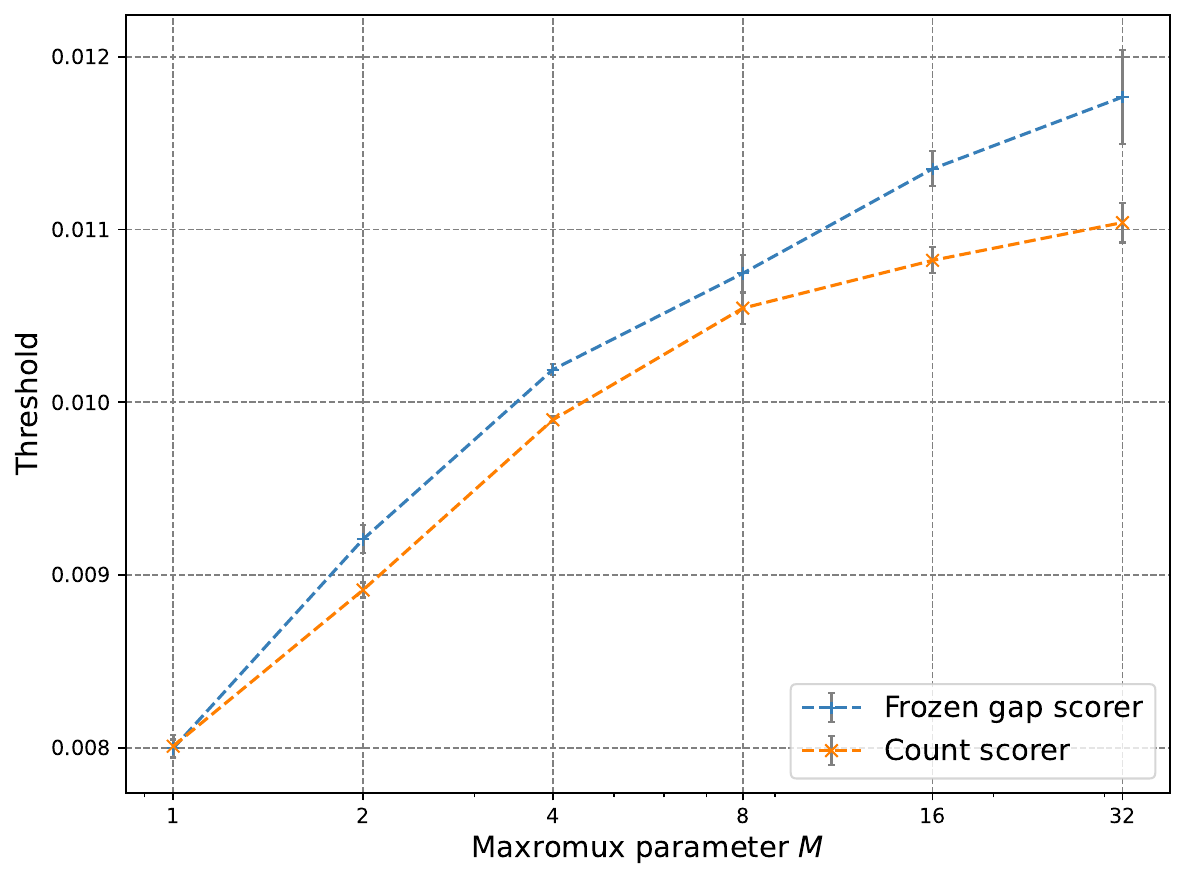} 
	\caption{(left) Thresholds as a function of the macromux parameter $M$ for different maximal brick sizes in the 6-ring fusion network. The error model is a mixed ray with both erasures and bitflips, each outcome is subject to erasure with rate $p_E$ and bitflip conditioned on no erasure with rate $p_P=p_E/10$. (right) Thresholds as a function of macromux overhead $M$ for both count and frozen gap scorer in the cuboctahedrally diced loopy diamond network. The error model consists of erasures and bitflips on fusions (some of which are correlated) in \textit{approximately} a 10 to 1 ratio. The frozen gap scorer improves the threshold by over $7\%$ at $M=32$ compared to the count scorer.}
	\label{figmmuxresults2}
\end{figure}

One can evaluate logical gaps between any two inequivalent logical sectors. Let $w_{\hat{l}}$ be the minimal weight correction obtained by decoding the Tanner graph with pseudo check nodes set to $\hat{l} = (l_1, \ldots, l_k) \in \mathbb{Z}_2^k$. Then the $2^k$ set of weights $w_{\hat{l}}$ for $\hat{l}\in \mathbb{Z}_2^k$ can be used to give a soft-output most-likely-error (MLE) decoder. 

One can recover the $i$th gap by considering two corrections. Let $\hat{l}^{\text{min}} = \arg\min\{|w_{\hat{l}}| ~|~ \hat{l} \in \mathbb{Z}_2^k\}$ be the minimal weight correction and $\hat{l}^{(i)} = \arg\min \{|w_{\hat{l}}| ~|~ \hat{l} \in \mathbb{Z}_2^k \text{ and } l_i \oplus l_i^{\text{min}} = 1\} $. Then $\Delta_i = |w_{\hat{l}^{\text{min}}} - w_{\hat{l}^{(i)}}|$.

\textbf{Frozen gap calculation.}
The frozen gap can be evaluated straightforwardly on the syndrome graph. For each disjoint boundary of each type (gap matching, or freeze), one can add a pseudo-vertex and connect to all degree-1 vertices on the respective boundary. For the initial recovery, each pseudo-vertex can take on either value in $\{0,1\}$. After freezing, the freeze pseudo-vertices are removed, and degree-1 vertices are also removed. The gap can be computed on this new configuration by varying the value of gap pseudo vertices in $\{0,1\}$ such that the overall configuration is mod-2 neutral. Fig.~\ref{figGapCalculation2} shows the syndrome graph for a 3x3x3 brick in the six-ring fusion network with freezing and gap boundaries. 

\section{Frozen gap scoring under mixed errors}\label{appMixedRay}

We have seen significant threshold improvement for loss-based, erasure-based and bitflip-based error models. In Fig.~\ref{figmmuxresults2} we show the threshold improvement for an error model that includes both erasures and bitflips. 

\section{Count vs frozen gap scoring and mixed errors}\label{appK6LO}

We have argued that the frozen gap scorer leads to better thresholds than count scoring, owing to its consideration of more fine-grained information about error configurations. In Fig.~\ref{figmmuxresults2} we show this improvement. While we have chosen a fusion network and error model motivated by linear optics, we remark that the benefit of frozen-gap scoring is generic.

\end{document}